\documentstyle[aps]{revtex}
\input epsf.tex

\def\apjl{{Ap.~J.~Lett.}}

\def\spose#1{\hbox to 0pt{#1\hss}}
\def\lta{\mathrel{\spose{\lower 3pt\hbox{$\mathchar"218$}}
     \raise 2.0pt\hbox{$\mathchar"13C$}}}
\def\gta{\mathrel{\spose{\lower 3pt\hbox{$\mathchar"218$}}
     \raise 2.0pt\hbox{$\mathchar"13E$}}}

\def\eg{{\it e.g.,\,}}
\def\ie{{\it i.e.,\,}}
\def\etal{{\it et al.\,}}

\def\be{\begin{equation}}
\def\ee{\end{equation}}
\def\bea{\begin{eqnarray}}
\def\eea{\end{eqnarray}}
\def\lbl{\label}
\def\hm{{\cal H}^3}
\def\xicx{\xi^{c}_\Phi}
\def\xic{\xi^{c}_\Phi({\bf x}, {\bf x}^\prime)}
\def\xico{\xi^{c}_\Phi({\bf x}, {\bf x})}
\def\xiu{\xi^{u}_\Phi({\bf x}, {\bf x}^\prime)}
\def\xiug{\xi^{u}_\Phi({\bf x}, \gamma {\bf x}^\prime)}

\def\xiur{\xi^{u}_\Phi(r)}
\def\xiurj{\xi^{u}_\Phi(r_j)}

\def\dT{\frac{\Delta T}{T}}

\def\tauls{\tau_{\sc ls}}
\def\taueq{\tau_{\sc eq}}
\def\cobedmr{{\sc cobe--dmr\,}}
\def\chiH{{\chi_{\!{}_{\rm H}}}}

\newcommand{\chix}[1]{{\chi_{\!{}_{\rm #1}}}}

\def\today{\ifcase\month\or
 January\or February\or March\or April\or May\or June\or
 July\or August\or September\or October\or November\or December\fi
 \space\number\day, \number\year}
\def\plotone#1{\centering \leavevmode
\epsfxsize= 0.6\columnwidth \epsfbox{#1}}
\def\plottwo#1#2{\centering \leavevmode
\epsfxsize=.6\columnwidth \epsfbox{#1} \hfil
\epsfxsize=.6\columnwidth \epsfbox{#2}}

\def\plottwosideR#1#2{\centering \leavevmode
\epsfxsize=.48\columnwidth \epsfbox{#1} \hfil
\epsfxsize=.48\columnwidth \epsfbox{#2}}

\begin{document}

\title {CMB Anisotropy in Compact Hyperbolic Universes II: COBE Maps
 and Limits} 
\author{J.Richard Bond, Dmitry Pogosyan and Tarun Souradeep}
\address{Canadian Institute for Theoretical Astrophysics,\\ University
of Toronto, ON M5S 3H8, Canada} \date{\today} \maketitle

\begin{abstract}
 The measurements of CMB anisotropy have opened up a window for
probing the global topology of the universe on length scales
comparable to,  and even beyond,  the Hubble radius. For compact topologies,
the two main effects on the CMB are: (1) the breaking of statistical
isotropy in characteristic patterns determined by the photon geodesic
structure of the manifold and (2) an infrared cutoff in the power
spectrum of perturbations imposed by the finite spatial extent.  We
calculate the CMB anisotropy in compact hyperbolic universe models
using the {\em regularized method of images}~\cite{us_texas,us_cwru}
described in detail in paper-I~\cite{paperA}, including 
the line-of-sight ``integrated Sachs-Wolfe'' effect~\cite{us_cwru}, as
well as the last-scattering surface terms~\cite{us_texas}. We calculate
the Bayesian probabilities for a selection of models by confronting
our theoretical pixel-pixel temperature correlation functions with the
\cobedmr data. Our results demonstrate that strong constraints on
compactness arise: if the universe is small compared to the `horizon'
size, correlations appear in the maps that are irreconcilable with the
observations. This conclusion is qualitatively insensitive to the
matter content of the universe, in particular, the presence of a
cosmological constant.  If the universe is of comparable size to the
'horizon', the likelihood function is very dependent upon orientation
of the manifold {\it wrt} the sky. While most orientations may be
strongly ruled out, it sometimes happens that for a specific
orientation the predicted correlation patterns are preferred over
those for the conventional infinite models. The full Bayesian analysis
we use is the most complete statistical test that can be done on the
COBE maps, taking into account all possible signals and their
variances in the theoretical skies, in particular the high degree of
anisotropic correlation that can exist. We also show that standard
visual measures for comparing theoretical predictions with the data
such as the isotropized power spectrum $C_\ell$ are not so useful in
small compact spaces because of enhanced cosmic variance associated
with the breakdown of statistical isotropy.
\end{abstract}

\section{Introduction}

The cosmic microwave background anisotropy is currently the most
promising observational probe of the global spatial structure of the
universe on length scales near to and even somewhat beyond the
`horizon' scale ($\sim c H_0^{-1}$).  As suggested by the concept of
inflation, this relatively smooth Hubble volume that we observe is
perhaps a tiny patch of an extremely inhomogeneous and complex spatial
manifold. The complexity could involve non-trivial topology (multiple
connectivity) on these ultra-large scales. Within a general program to
address the observability of such a diverse global structure, a more
well defined and tractable path would be to restrict oneself to spaces
of uniform curvature (locally homogeneous and isotropic FRW models)
but with non-trivial topology; in particular, compact spaces which
have additional theoretical
motivation~\cite{ell71,sok_shv74,lac_lum95,got80,cor9596}. For
Euclidean (uniform zero curvature) or hyperbolic (uniform negative
curvature) geometry, compactness necessarily implies non-trivial
topology.  Much recent astrophysical data suggest the cosmological
density parameter in matter is subcritical~\cite{opencase}, $\Omega_m < 1$.
Recently the presence of a significant cosmological constant (or, more
generally, an exotic smooth component of matter) has been indicated by
the high redshift supernova searches~\cite{SNres} and in studies
combining large and intermediate angle CMB anistropy data with
observations of cluster abundances and large
scale galaxy clustering~\cite{BJroysoc}. If this component
does not compensate for the deficit from unity,
$\Omega_0=\Omega_m+\Omega_\Lambda<1$, this would imply a hyperbolic
spatial geometry for the universe; the additional requirement of
compactness then ushers into consideration {\em topologically compact
hyperbolic (CH) universes}, a field much richer in possibilities
than the compact spaces with flat geometry when $\Omega_0$ is {\it
exactly} unity.

In a universe with non-trivial global spatial topology, the multiple
connectivity of the space could lead to observable characteristic
angular correlation patterns in the CMB anisotropy arising directly
from multiple imaging of the source terms that give rise to the
anisotropy in the CMB. Moreover, the modified structure of the
eigenmodes in such spaces implies that angular correlations would
differ from the predictions in the simply connected space with identical
geometry (the latter plays the role of the {\em universal cover} of
the multiply connected space), even in the absence of multiple imaging
of the sources.  In particular, compact universes cannot support modes
whose characteristic length scale exceeds the linear size of the
space; consequently, the inferred power of fluctuations in compact
models at large scales would appear suppressed relative to the power
on the universal covering space. A more subtle effect is the angular
dependence of the theoretical temperature variance which reflects
generic inhomogeneity in the topologically compact spaces.

Paper-I~\cite{paperA} describes the {\em regularized method of
images}, a general technique that we developed~\cite{us_texas,us_cwru}
for computing the spatial correlations in a universe with non-trivial
topology. In this paper we shall apply the method to calculate the
angular correlation of CMB anisotropies in CH universe models. The
angular correlation between the CMB temperature fluctuations in two
directions in the sky completely encodes the CMB anisotropy
predictions of any model that postulates Gaussian primordial
fluctuations. Using the full correlation function information on the
\cobedmr data, we have obtained limits on the size of flat torus
universe models that are a factor of two sharper than obtained from
the angular power spectrum alone~\cite{us_torus}.  The main result was
that the volume of the compact universe is constrained to be
comparable to or larger than that of the observable universe. In
\cite{us_texas}, we proposed on the basis of simple arguments that
compact universes with hyperbolic geometry should be expected to
respect similar constraints. We have now carried out the full Bayesian
analysis on accurate CMB correlation predictions in a large selection
of CH universes and can demonstrate that essential features of our
general constraint is borne out. In this paper we describe the details
of our CMB correlation computation, highlight the general correlation
features in the compact spaces, describe the method of comparison to
data and present the resulting constraints from \cobedmr using the
examples of a large and a small CH universe. A compilation of our
constraints on the flat torus models and a large selection of CH
models will be presented in a separate publication~\cite{us_prl}.

The outline of this paper is as follows: In $\S$\ref{sec_cmb} we
recapitulate that the angular correlation function of CMB anisotropy
on large angular scales, that predominantly arises through both
surface and integrated Sachs-Wolfe effects, can be related to spatial
correlations of the gravitational potential on the three-space
hypersurface at the epoch of last scattering. This is a useful
simplification in terms of computational costs for calculating the CMB
correlation in compact spaces using our method. Although we restrict
our calculations here to large angular scales,
Appendix~\ref{sec_cthetafull} discusses the implementation of the
method of images to calculate the CMB anisotropies at smaller angular
scales. In $\S$\ref{sec_CHM}, we describe the computation of the
angular correlation function of CMB anisotropy in CH models. The
section also includes a quick review of some useful notions about
compact spaces and the main result describing the regularized method
of images from paper-I~\cite{paperA}. 

In $\S$\ref{sec_cmbcorr}, we discuss some of the typical correlation
features in the CMB anisotropy that arise in compact universe models.
We show that their origin is more readily understood by viewing the
compact space as a tessellation of $\hm$ by the finite domains.  In
$\S$\ref{sec_prob}, we present our results of full Bayesian
probability analyses of large angle CMB anisotropy predictions for two
CH models using the four year \cobedmr data. In
Appendix~\ref{sec_cthetafull}, we show how to go beyond the
Sachs-Wolfe effects to treat all aspects of CMB anisotropy using the
method of images in compact spaces for high resolution CMB
experiments. In Appendix~\ref{sec_noCl}, we demonstrate that a
byproduct of the statistical anisotropy of the CMB inherent in compact
universe models is a considerably enhanced cosmic variance in the
(isotropized) angular power spectrum $C_\ell$ which completely characterizes the
non-compact Gaussian models. This emphasizes that the pattern
recognition aspect of the complete Bayesian testing of a model is
essential to get the best constraints on allowed size of the compact
space.

\section{CMB anisotropy}
\lbl{sec_cmb}

In the standard picture, the CMB that we observe is a Planckian
distribution of relic photons which decoupled from matter at a
redshift $\approx 1100$. These photons have freely propagated over a
distance $ R_{\sc ls}$, comparable to the ``horizon'' size, a function
of cosmological parameters. In a non-flat model the other length scale is
curvature radius $d_c$ given by $(c/H_0)/\sqrt{|1-\Omega_0|}$.
In a matter dominated ($\Omega_0=\Omega_m$) cosmology,
$R_{\sc ls}\approx 2 d_c\, {\rm arctanh} \sqrt{1-\Omega_0}$.
For the adiabatic fluctuations we consider here, the dominant
contribution to the anisotropy in the CMB temperature measured with
wide-angle beams ($\theta_{\sc fwhm} \gta 2^\circ \Omega_0^{1/2}$)
comes from the cosmological metric perturbations through the
Sachs-Wolfe effect. Although in this work we restrict our attention to
large beam size and the Sachs-Wolfe effect, in
Appendix~\ref{sec_cthetafull} we show how effects which contribute to
the CMB anisotropy at finer resolution can also be incorporated in the
method of images.

Adiabatic cosmological metric perturbations can be expressed in terms
of a scalar gravitational potential $\Phi({\bf x},\tau)$.  The
dynamical equation for $\Phi({\bf x},\tau)$ allows for separation of
the spatial and temporal dependence in the linear regime,\footnote{At
the scales appropriate to CMB anisotropies, damping effects on $\Phi$
can be neglected.}  $\Phi({\bf x},\tau) = (F(\tau) + E(\tau))
\Phi({\bf x})$, where $\Phi({\bf x})$ is the field configuration on
the three-hypersurface of constant time, whose amplitude is determined
by the physics of the early universe. We use as our time variable 
dimensionless conformal time $\tau$,  expressed in units of the
curvature radius $d_c$. Time dependence of the potential
at the matter dominated stage is described by the 'growing' mode
$F(\tau)$ and the decaying mode $E(\tau)$. In terms of the usual
growth factor $D(t)$ for linear density perturbations, $F=D/a$, where
$a$ is the scale factor. The relative amplitude of the modes is
determined by the matching condition at the moment $\taueq \approx
0.004 h^{-1}\sqrt{1-\Omega_0}/\Omega_0$ of transition from
extremely-relativistic to non-relativistic domination of the energy
density. This gives $E(\taueq)\approx F(\taueq)/9$.  

In this paper, we concentrate our study on open matter-dominated
models with zero cosmological constant, $\Omega_0=\Omega_m<1$, for
which the growing mode evolves as~\cite{mukh_92}
\begin{equation}
F(\tau) =  \frac{ 5\sinh\tau(\sinh\tau-3\tau)+20(\cosh\tau -1)}
{(\cosh\tau -1)^{3}}
\label{Feta}
\end{equation}
in the matter dominated phase, $\tau >\taueq $.  A non-zero
cosmological constant can be trivially incorporated in our analysis by
using the appropriate solution for $F(\tau)$.

We write the Sachs-Wolfe formula for the CMB temperature fluctuation,
$\Delta T(\hat q)$, in a direction $\hat q$, in terms of the growing
mode $\Phi_{\sc ls}({\bf x})$ of the potential at the
three-hypersurface of constant time $\tau=\tauls$, when the last
scattering of CMB photons took place, $\Phi_{\sc ls}({\bf
x})=F(\tauls)\Phi({\bf x})$
\footnote{A subtle aspect of the Sachs-Wolfe effect at large angular
scales is that only the growing mode $F(\tauls)\Phi({\bf x})$ and {\it
not} the total gravitational potential at last-scattering $\Phi({\bf
x},\tauls)= (F(\tauls)+E(\tauls))\Phi({\bf x})$ contributes to the
effect.  The distinction is important for models with small value of
$\Omega_m$ in which the time difference between the transition to
matter domination at $\taueq $ and the last scattering of photons at
$\tauls$ is not large. With this important caveat, we shall still loosely call
$\Phi_{\sc ls}({\bf x})$ the potential on the last-scattering
hypersurface.}

\begin{equation}
\dT(\hat q) = \frac{1}{3} \Phi_{\sc ls}(\hat q\chiH) +
2 \int_{0}^\chiH d\chi f(\chi)
 \Phi_{\sc ls}(\hat q\chi)\,, \quad 
 f(\chi)=\frac{1}{F(\tauls)}\frac{d}{d\tau} F(\tau)\bigg|_{\tau=\tauls+\chiH-\chi}\,,
\lbl{dTSW}
\end{equation}
where $\chi$ is the affine parameter along the photon path from
$\chi=0$ at the observer position to $\chiH=R_{\sc ls}/d_c$. The first
term is called the {\em surface} or ``naive'' Sachs-Wolfe effect
(NSW). The second term, which is nonzero only if $\Phi$ varies with
time between $\tauls$ and now, is the {\em integrated} Sachs-Wolfe
effect (ISW). The angular correlation between the CMB temperature
fluctuations in two directions in the sky is then given by

\begin{eqnarray}
 C(\hat q,\hat q^\prime)\equiv
\left\langle\dT(\hat q)\dT(\hat q^\prime)\right\rangle
&&{}= \frac{1}{9} 
\langle\Phi_{\sc ls}(\hat q\chiH)\Phi_{\sc ls}(\hat q^\prime\chiH)\rangle
\nonumber\\
&&{} +\frac{2}{3}\int_{0}^\chiH d\chi~f(\chi)~\left[
\langle\Phi_{\sc ls}(\hat q\chi)\Phi_{\sc ls}(\hat q^\prime\chiH)\rangle 
+\langle\Phi_{\sc ls}(\hat q^\prime\chi)\Phi_{\sc ls}(\hat q\chiH)\rangle\right]
\nonumber\\
&&{} +4\int_{0}^\chiH d\chix{1} ~f(\chix{1})~
 \int_{0}^\chiH  d\chix{2}~f(\chix{2})~
\langle\Phi_{\sc ls}(\hat q\chix{1})\Phi_{\sc ls}(\hat q^\prime\chix{2})\rangle\,.
\lbl{cthetaSW}
\end{eqnarray}

The main point to be noted is that $C(\hat q,\hat q^\prime)$ depends
on the spatial two point correlation function, $\xi_\Phi \equiv
\langle\Phi_{\sc ls}({\bf x})\Phi_{\sc ls}({\bf x^\prime})\rangle $ of
$\Phi$ on the three-hypersurface of last scattering. This is due to
the fact that the equation of motion for $\Phi$ allows a separation of
spatial and temporal dependence.  As in eq.({\ref{dTSW}), the
Sachs-Wolfe contribution to the correlation between temperature
fluctuations in two directions in the sky can be split into three
terms : 1) The surface term (NSW) which depends on the correlation
between $\Phi$ at the two points on the SLS; 2) the interference part
correlating the value of $\Phi$ at the points along one line of sight
to the value at the SLS of the second line of sight; and, 3) an integral
part which contains correlations between $\Phi$ at points on 
one line of sight with those on the other. The last two terms
constitute the ISW effect.  If one considers the zero-lag correlation
(the two lines of sight are identical), then the following holds: the
first and third terms are positive definite, whereas the interference
term comes in with a negative sign because $dF(\tau)/d\tau$ is
negative in the models that we consider here.

In $\hm$, the global isotropy of the space implies that the two point
correlation function $C(\hat q,\hat q^\prime) \equiv C(\theta)$, where
$\cos\theta=\hat q\cdot\hat q^\prime$, and the CMB anisotropy can be
described equally well in terms of its angular power spectrum ${\cal
C}_\ell$, defined by
\begin{equation}
C(\theta) = \sum_{\ell} \frac{\ell +1/2}{\ell(\ell +1)}~{\cal C}_\ell
P_\ell(\cos\theta) W_\ell \, , 
\label{Ctheta_Cl}
\end{equation}
where the $P_\ell$ are Legendre polynomials and $W_\ell$ encodes
details of the experimental configuration, such as finite
beamwidth. For \cobedmr, $W_\ell =B_\ell^2$, where $B_\ell$ is the
beam, including a (sphericalized) approximation to finite pixelization
effects.  In this paper, we use the experimentally-determined $W_\ell$
for \cobedmr, but, to set the scale, we note that a Gaussian $B_\ell
\sim \exp\left[-\ell(\ell+1)\sigma^2_{\rm beam}/2\right]$ fit to the
\cobedmr beam (including pixelization effects) gives $\sigma_{\rm
beam}^{-1}\approx 17.5$.~\footnote{ When we speak about CMB anisotropy
at ``large angular scales'' we always refer to the beam size
$\sigma_{\rm beam}$ and not the angular separation between the lines
of sight, $\theta$. This distinction is blurred in usual simply
connected models. There, for example, the SW effect dominates both
wide beam experiments such as \cobedmr and also the correlation at
large separation $\theta$.  In contrast, the distinction is important
for compact spaces where points widely separated in angle may be
physically close (see $\S$\ref{sec_cmbcorr}), resulting in the
correlation between them being dominated not by SW but by
short-distance effects, including the Doppler effect.  It should also
be noted that since compact universe models generically violate global
isotropy, the decomposition (\ref{Ctheta_Cl}) is not valid in those
cases (the angular power in a multipole $\ell$ is not evenly
distributed in the azimuthal levels, $m$).}  

The angular power spectrum of the CMB anisotropy on large angular
scales arises mainly from the Sachs-Wolfe effect and is shown in
Figure~\ref{fig:clisw}. The ${\cal C}_\ell$ contribution of the NSW
term is plotted separately for comparison. The NSW contribution is
suppressed at angular scales larger than the curvature scale due to
focusing of geodesics in the hyperbolic geometry~\cite{wil82} and
asymptotes to a constant value at small angular scales. The ISW
contribution to ${\cal C}_\ell$ falls off roughly a factor of
$\ell^{-1}$ faster than the NSW contribution.  At small values of
$\Omega_0$, the ISW contribution dominates at large angular scales.

\begin{figure}[tbh]
\plotone {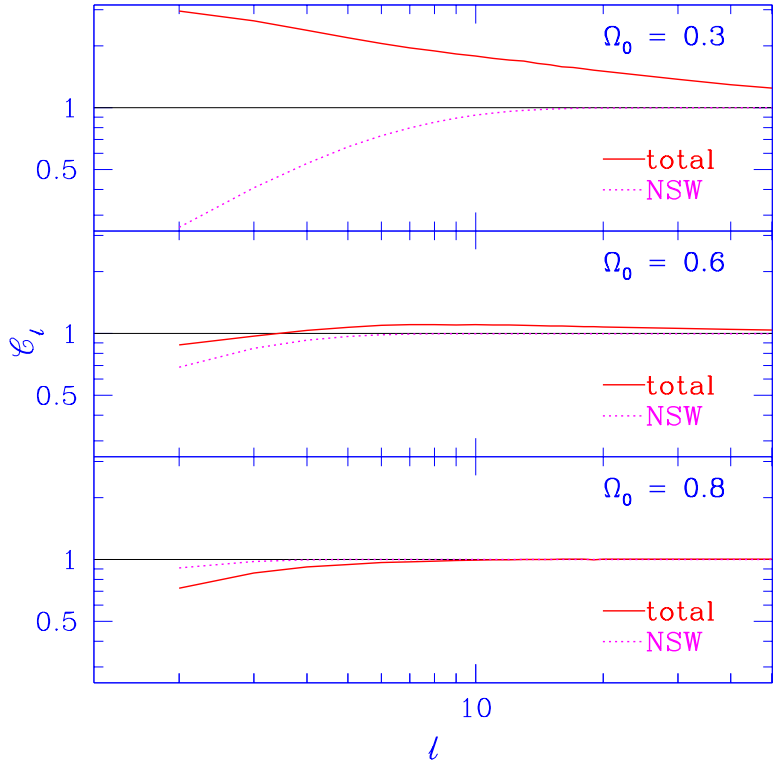}
\caption{ The solid curves show the angular power spectrum ${\cal
C}_\ell$ from the Sachs-Wolfe effect in the three infinite $\hm$
universes with differing $\Omega_0$. Also plotted as a dotted curve is
the NSW contribution in each case. The curves are normalized such that
the ${\cal C}_\ell$ from the NSW contribution goes to unity at large
$\ell$. As $\Omega_0$ increases towards unity, the relative ISW
contribution diminishes and affects smaller values of $\ell$.  While
the ISW contribution is positive for $\Omega_0=0.3$ and
$\Omega_0=0.6$, it is negative for $\Omega_0=0.8$.  }
\label{fig:clisw}
\end{figure}

Figure~\ref{fig:isw_kdc} illustrates some general features of the
distribution of power in $k$-space for the low multipoles of the
Sachs-Wolfe CMB anisotropy, using the fourth multipole as an example.
While the NSW contribution is always positive, the ISW contribution
can be negative due to the interference term.  The positive ISW
contribution comes from larger values of $k$ than for the NSW effect. 
Moreover, the small $k$ contribution of the positive
NSW effect is countered by the negative interference term in the ISW
contribution (the second term in eq.~(\ref{cthetaSW})). Note the
remarkable cancellation of the NSW contribution for $\Omega_0=0.3$.
The relative contribution of the ISW effect decreases as $\Omega_0\to
1$. The presence of the ISW contribution tends to relax the constraints on
the size of a compact universe that can be directly inferred from the
suppression of the low multipoles of the CMB anisotropy~\cite{cor98}.

\begin{figure}[tbh]
\plotone{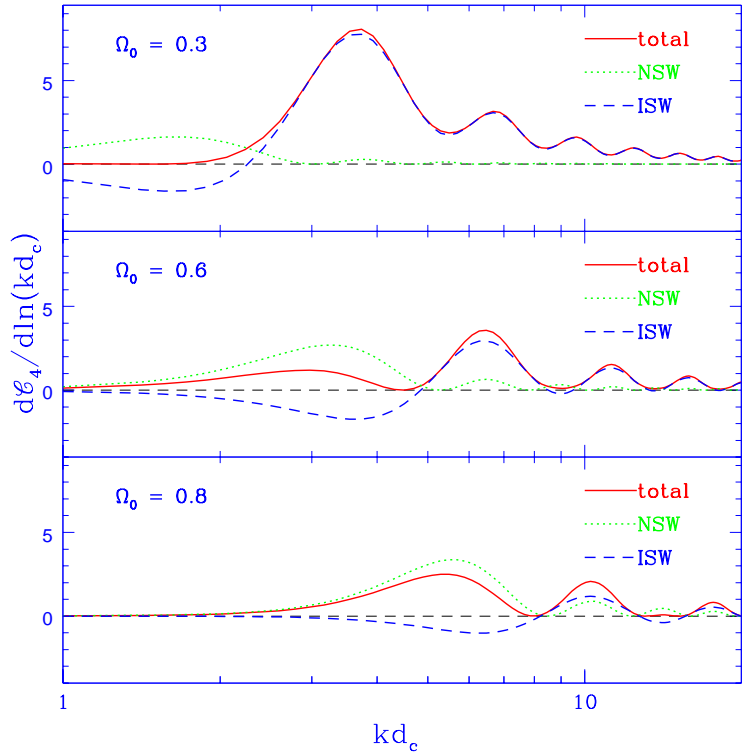}
\caption{The $k$-space dependence of the integrand for the fourth
multipole ${\cal C}_4$ from the Sachs-Wolfe effect in 3 
infinite $\hm$ universes characterized by the values of $\Omega_0$ shown.
In each case the dotted and dashed curves are the NSW and ISW
contributions, respectively. The ISW curve includes the NSW-ISW interference
term as well.}
\label{fig:isw_kdc}
\end{figure}

\section{CMB anisotropy in Compact hyperbolic manifolds}
\lbl{sec_CHM}

\subsection {Brief review of compact spaces}

We briefly recapitulate a few basic notions about compact universes
that we discussed in paper-I~\cite{paperA}. A compact cosmological
model can be constructed by identifying points on the standard
infinite flat or hyperbolic FRW spaces by the action of a suitable
discrete subgroup of motions, $\Gamma$, of the full {\em isometry
group}, $G$, of the FRW space.  (The isometry group $G$ is the group
of motions which preserves the distances between points, \ie leaves
the metric unchanged). The infinite FRW spatial hypersurface is the
{\em universal cover}, ${\cal M}^u$, tiled by copies of the compact
space ${\cal M}$. The compact space for a given location of the
observer is most appropriately represented as the {\em Dirichlet
domain} with the observer at its {\em basepoint}.  Any point ${{\bf
x}}$ of the compact space has an image ${{\bf x}}_i = \gamma_i {{\bf
x}}$ in each copy of the Dirichlet domain on the universal cover,
where $\gamma_i \in \Gamma$.  The tiling of the universal cover with
Dirichlet domains is a Voronoi tessellation, (a familiar concept in
cosmology often used in modeling the large scale structure in the
universe), with the seeds being the basepoint and its images. By
construction a Dirichlet domain represents the compact space as a {\em
convex polyhedron} with even number of faces identified pairwise under
$\Gamma$. In cosmology, the Dirichlet domain constructed
around the observer represents the universe as `seen' by the observer
and it proves useful in this context to define the {\em outradius},
$R_>$, the radii of the circumscribing sphere (smallest sphere around
the observer which encloses the Dirichlet domain) and, the {\em
inradius}, $R_<$, the radii of the inscribed sphere (largest sphere
around the observer which can be enclosed within the Dirichlet domain)
of the Dirichlet domain~\cite{sok_shv74}. Note that $R_>$ and $R_<$
are specific to the location of the observer within the compact space
since the Dirichlet domains around different observers are not
necessarily identical.  An observer-independent (and Dirichlet-domain-independent) 
linear measure of the size of the compact space is given by the {\em
diameter} of the space, $d_{\!\cal M} \equiv \sup_{x,y\in{\cal
M}}d(x,y)$~\cite{ber80,chav84}, \ie the maximum separation between two
points in the compact space.

For cosmological CH models, ${\cal M}^u \equiv \hm$, the
three-dimensional hyperbolic (uniform negative curvature) manifold.
$\hm$ can be viewed as a hyperbolic section embedded in four
dimensional flat Lorentzian space. The isometry group of $\hm$ is the
group of rotations in the four space -- the proper Lorentz group,
$SO(3,1)$. A CH manifold is then completely
described by a discrete subgroup, $\Gamma$, of the proper Lorentz
group, $SO(3,1)$. The Geometry Centre at the University of Minnesota
has a large census of CH manifolds and public domain software
SnapPea~\cite{Minn}. We have adapted this software to tile $\hm$ under
a given topology using a set of generators of $\Gamma$.  The tiling
routine uses the generator product method and ensures that all
distinct tiles within a specified tiling radius are obtained.

A CH manifold, ${\cal M}$, is characterized by a dimensionless number,
${\cal V}_{\!\cal M}\equiv V_{\!\cal M}/d_c^3$, where $V_{\!\cal M}$
is the volume of the space and $d_c$ is the curvature radius
\cite{Thur7984}. There are a countably infinite number of CH manifolds
with no upper bound on ${\cal V}_{\!\cal M}$. The theoretical lower
bound stands at ${\cal V}_{\!\cal M} \ge 0.167$~\cite{gab_mey96}. The
smallest CH manifold discovered so far has ${\cal V}_{\!\cal M}
=0.94$~\cite{smallestCH}.  The Minnesota census lists several
thousands of these manifolds with ${\cal V}_{\!\cal M}$ up to $\sim
7$. In the cosmological context, the physical size of the curvature
radius $d_c$ is determined by the density parameter and the Hubble
constant $H_0$: $d_c=(c/H_0)/\sqrt{1-\Omega_0}$. The physical volume
of the CH manifold with a given topology, \ie a fixed value of
$V_{\!\cal M}/d_c^3$, is smaller for smaller values of $\Omega_0$.

\subsection {Computing spatial correlation functions}

Here we summarize the main result of our regularized method of images
that is discussed in detail in paper-I~\cite{paperA}.  The correlation
function on a compact space (and more generally, any non-simply
connected space), ${\cal M}={\cal M}^u/\Gamma$, can be expressed as a
regularized sum over the correlation function on its universal cover,
${\cal M}^u$, calculated between ${\bf x}$ and the images $\gamma{\bf
x^\prime}$ ($\gamma\in\Gamma$) of ${\bf x^\prime}$:
\begin{eqnarray}
\xic &=& \widetilde{\sum_{\gamma\in\Gamma}} \xiug \nonumber \\
&=& \sum_{\gamma\in\Gamma} \xiug -\frac{1}{V_{\!\cal
M}}\int_{{\cal M}^u}d{\bf x^\prime}~\xiu \,. 
\lbl{moi1}
\end{eqnarray}

The local isotropy and homogeneity of $\hm$ implies $\xiu$ depends
only on the proper distance, $r\equiv d({\bf x},{\bf x^\prime})$,
between the points ${\bf x}$ and ${\bf x^\prime}$.  The eigenfunctions
on the universal cover are of course well known for all homogeneous
and isotropic models~\cite{har67}. Consequently $\xiu$ can be obtained
using its eigenmode expansion.  The initial power spectrum $P_\Phi(k)$
is believed to be dictated by an early universe scenario for the
generation of primordial perturbations. We assume that the initial
perturbations are generated by quantum vacuum fluctuations during
inflation. This leads to
\begin{equation} 
\xiu \equiv \xiur = \int_0^\infty
\frac{d\beta~\beta}{(\beta^2 +1)} ~\frac{\sin(\beta r)}{\beta \sinh
r}~~{\cal P}_\Phi(\beta) \, , \lbl{xiur}
\end{equation}
where $\beta \equiv \sqrt{(k d_c)^2 -1}$ and ${\cal
P}_\Phi(\beta)\equiv \beta(\beta^2+1) P_\Phi(k)/(2\pi^2)$.

In the simplest inflationary models, ${\cal P}_\Phi$, the power per
logarithmic interval of $k$ is approximately constant in the
subcurvature sector, defined by $k d_c > 1$.  This is the
generalization of the Harrison-Zeldovich spectrum in spatially flat
models to hyperbolic spaces~\cite{lyt_stew90,rat_peeb94}.  Sub-horizon
vacuum fluctuations during inflation are not expected to generate
supercurvature modes, those with $k d_c <1$, which is why they are not
included in eq.(\ref{xiur}). Indeed, since $H^2 > 1/(a d_c)^2$, for
modes with $k d_c < 1$ we always have $k/(aH) < 1$ so inflation by
itself does not provide a causal mechanism for their
excitation. Moreover, the lowest non-zero eigenvalue in
compact spaces, $k_1>0$,  provides an infra-red cutoff in the spectrum which can
be large enough in many CH spaces to exclude the supercurvature sector
entirely ($k_1 d_c > 1$).  (See ~\cite{paperA}.)  Even if the space
does support supercurvature modes, some physical mechanism needs to be
invoked to excite them, \eg as a byproduct of the creation of the
compact space itself, but which could be accompanied by complex
nonperturbative structure as well.  To have quantitative predictions
for $P_\Phi(k)$ would require addressing this possibility in a full
quantum cosmological context.  We note that our main conclusions
regarding peculiar correlation features in the CMB anisotropy (see
$\S$~\ref{sec_cmbcorr}) would qualitatively hold even in the presence
of supercurvature modes.

Although equation (\ref{moi1}) encodes the basic formula for
calculating the correlation function, it is not numerically
implementable as is. Both the sum and the integral in equation
(\ref{moi1}) are divergent and the difference needs to be taken as a
limiting process of summation of images and integration up to a finite
distance $r_*$:
\begin{equation}
\xic = \lim_{r_*\to\infty} \left[ \sum_{r_j < r_*} \xiurj - 
\frac{4\pi}{V_{\!\cal M}} \int_0^{r_*} dr ~\sinh^2r ~\xiur\right],~~~~
r_j = d({\bf x}, \gamma_j {\bf x}^\prime)\, . 
\lbl{moi_final}
\end{equation}
The volume element in the integral is the one appropriate for $\hm$. Numerically
we have found it suffices to evaluate the above expression up to $r_*$
about $4$ to $5$ times the domain size $R_>$ to obtain a convergent
result for $\xic$.

\subsection{Computing the CMB correlation}
\label{ssec_CH_cmbcalc} 
For Gaussian perturbations, the angular correlation function, $C(\hat
q,\hat q^\prime)$, completely encodes the CMB anisotropy predictions
of a model. To make maps, the celestial sphere is discretized into
$N_p$ pixels labeled by $p$. $N_p$ is determined by the angular
resolution, $\sim 7^\circ$ degrees for \cobedmr. The \cobedmr maps had
$(2.6^\circ)^2$ pixels, corresponding to $N_p=6144$, though
compressing the data into $N_p=1536$ pixels leads to no information
loss. We use this number of pixels in \S~\ref{sec_prob}, where we
confront our models with the \cobedmr observations using Bayesian
statistical methods, and also in the maps shown in this paper. 

For Gaussian statistics, the pixelized theoretical maps are fully
determined by the $N_p\times N_p$ pixel-pixel correlation matrix,
$C_{Tpp^\prime}\equiv C(\hat q_p,\hat q_{p^\prime})$.  The expression
for $C(\hat q,\hat q^\prime)$ in eq.~(\ref{cthetaSW}) involves an
integral along the line of sight from the observer to the surface of
last scattering -- the integrated Sachs--Wolfe term (ISW). We find
that a simple integration rule using $N_L$ points along each line of
sight gives accurate results when the points are spaced at equal
increments $\Delta F $ of $F(\tau)$. Consequently, to get the full
$C_{Tpp^\prime}$ matrix we need to evaluate the correlation function,
$\xic$, between $N_p(N_p+1) N_L^2/2$ pairs of points on the
constant-time hypersurface of last scattering:
\begin{equation}
 C_{Tpp^\prime} = \frac{1}{9} \xicx(\chiH \hat q_p,\chiH \hat
q_{p^\prime}) + \frac{2}{3}~\frac{\Delta F}{N_L}\sum_{i=1}^{N_L}
w_i\left[ \xicx(\chiH \hat q_p, \chi_i \hat q_{p^\prime}) +\xicx(\chi_i
\hat q_{p^\prime}, \chiH\hat q_p) \right] + 4~\left(\frac{\Delta F}{N_L}
\right)^2
\sum_{i=1}^{N_L}\sum_{j=1}^{N_L} w_{ij}\xicx(\chi_i \hat q_{p^\prime},
\chi_j\hat q_p) 
\label{ctppSW}
\end{equation}
where $w_i$ and $w_{ij}$ are ${\cal O}(1)$ coefficients that depend on
the specific difference scheme used.

To calculate $C_{Tpp^\prime}$ in the standard infinite open
cosmological models with this real-space integration to an accuracy
comparable to that of the traditional evaluation in \mbox{$k$--space},
$N_L \sim 10$ is sufficient. Moreover, the real space
integration is much faster in terms of CPU time. For the method to
remain accurate in compact models, $N_L$ should, of course, exceed the
number of the times a typical photon path crosses the compact
Universe.  We found $N_L \sim 10$ is still enough for the models we
have analyzed so far.

\section{Correlation features  in  the CMB anisotropy }
\label{sec_cmbcorr}

In standard cosmological models based on a topologically trivial
space such as $\hm$, the observed CMB photons have propagated along
radial geodesics from a $2$-sphere of radius $R_{\sc ls}$ (that we
refer to as the sphere of last scattering, SLS) centered on the
observer. The same picture also applies to CH models when the space is
viewed as a tessellation of the universal cover tiled by the Dirichlet
domain with the observer at the basepoint.  If $V_{\sc LS}$, the
volume of the SLS, is much larger that that of the compact space, the
photons propagate through a lattice of identical domains. As a
consequence, strong correlations build up between CMB temperature
fluctuations observed in widely separated directions.  The correlation
function $C (\hat q, \hat q^\prime)$ is anisotropic and contains
characteristic patterns determined by the photon geodesic structure of
the compact manifold. These correlations persist even in CH models
whose Dirichlet domains are comparable to or slightly bigger than the
SLS. This is the key difference from the standard models where $C
(\hat q, \hat q^\prime) $ depends only on the angle between $\hat q$
and $\hat q^\prime$ and generally falls off with angular separation.

The pattern of strong correlations is directly linked to the way the
points (on the universal cover ${\cal M}^u$) enclosed by the SLS are
equivalent under the topological identification $\Gamma$. Consider a
set of points $S$ on the universal cover. Then for all $\gamma \in
\Gamma$ the pair of subsets $\gamma S\cap S$ and $S \cap\gamma^{-1} S$
are equivalent under $\Gamma$, \ie either contain the same points of
the compact space ${\cal M}$ or they are both empty. (Obviously, the
isometry $\gamma^{-1}\in\Gamma$ maps the first set onto the second.)
If $S$ consists of the points on the SLS, a non-empty set $S\cap\gamma
S $ would be a circle on the SLS which is pointwise identical to
another circle $S\cap\gamma^{-1}S$ on the SLS.  If the CMB anisotropy
is dominated by the surface term, then the CMB temperature along one
circle is expected to be identical to that along the other
circle. This effect was first understood and much emphasized in the
works byCornish, Starkman and Spergel~\cite{circles}.  The
necessary and sufficient condition for the existence of matched
circles (non empty subsets $S\cap\gamma S$) on the SLS is that $R_{\sc
ls}\ge R_<$, \ie the SLS is not completely enclosed
within the Dirichlet domain. We discuss the CMB correlations in our models that
arise from this kind of identification in $\S$~\ref{ssec_cmbcorr_nsw}.

\begin{figure}[h]
\plotone{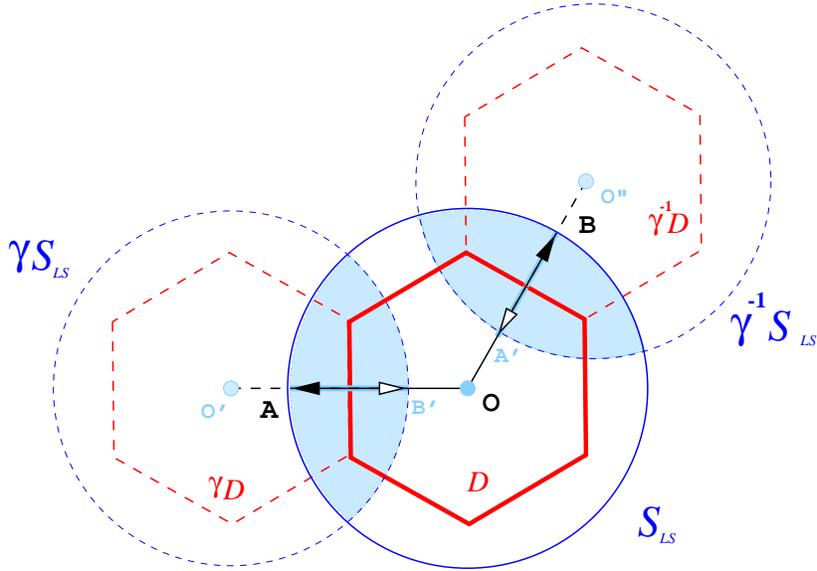}
\caption{ The origin of CMB correlation patterns in a multiply
connected universe is illustrated. The compact space is represented by
the (hexagonal) Dirichlet domain, $D$, around the observer, $O$, which
tiles the universal cover. The tiles $\gamma D$ and $\gamma^{-1}D $ 
under one of the face translations $\gamma\in\Gamma$ and its inverse
are shown.  The points $O^\prime$ and $O^{\prime\prime}$ are the
images of the observer, $O$, under $\gamma$ and $\gamma^{-1}$,
respectively. The point $B^\prime$ is the $\gamma$-translate of $B$
and the point $A^\prime$ is the $\gamma^{-1}$ translate of $B$.  The
solid circle $S_{LS}$ represents the sphere of last scattering
(SLS) on the universal cover; the dashed circles, labeled $\gamma
S_{LS}$ and $\gamma^{-1} S_{LS}$, are the translates of SLS under
$\gamma$ and $\gamma^{-1}$ , respectively. The intersections of the
sphere $S_{LS}$ with $\gamma S_{LS}$ and $\gamma^{-1}S_{LS}$ create a
matched pair of circles on the SLS. Similarly, the shaded lens-shaped
regions within the $SLS$ consist of identical sets of points. Consider
the lines of sight $OA$ and $OB$ directed towards the centers of the
matched circles. The ray $O^\prime B^\prime$ is the image of $OB$
under $\gamma$ and the ray $O^{\prime\prime}A^\prime$ is the image of
$OA$ under $\gamma^{-1}$. By the arguments outlined in the text, the
segment $AB^\prime = OA\cap OB^\prime$ is identical to $A^\prime B =
OB \cap OA^\prime$.}
\label{fig:top}
\end{figure}

The CMB anisotropy has contributions from the entire line of sight and
it is useful to study the subsets of identical points enclosed by the
SLS.  If $S$ is the set of all points enclosed by the SLS, then the
pairs of subsets of $S$ that are identical are lens-shaped regions
created by the intersection of two balls (See Figure~\ref{fig:top}).
Again the necessary and sufficient condition for the existence of such
identified regions within the SLS is $R_{\sc ls}\ge R_<$.  Further if
the SLS does not enclose more than one Dirichlet domain, these
lens-like regions are in the direction of the faces and are adjacent
to the SLS.  (We find from our analysis of the \cobedmr constraints
that viable CH models have volumes comparable to or more than that
enclosed within the SLS. See $\S$~\ref{sec_prob}.) In these cases it
is clear that correlation patterns are built from points close to the
SLS and directly reflect the shape of polyhedral Dirichlet domain.

The ISW-CMB correlation depends on the correlations between the points
lying on the two (radial) lines of sight from the observer to the SLS.
Hence it is instructive to study whether two different lines of sight
(radial lines) in the SLS contain a set of identical points. Consider
the set of points $L_1$ and $L_2$ along two distinct lines of
sight. Then for all $\gamma\in\Gamma$, the subset $L_1\cap \gamma^{-1}
L_2$ of $L_1$, if non-empty, would contain the same points as the
subset $\gamma L_1\cap L_2$ of $L_2$. The necessary and sufficient
condition for the existence of such pairs of lines of sight is again
$R_{\sc ls}\ge R_<$. It is possible to show that there are infinite
pairs of lines of sights which share at least one common point, but
what is more interesting and relevant are the pairs where the
identical subsets contain a segment. It is a straightforward exercise
to verify that every pair of lines of sight pointing towards centers
of matched circles discussed above must contain a segment consisting
of identical points (See Figure~\ref{fig:top}). As we shall discuss in
$\S$~\ref{ssec_cmbcorr_isw}, this result is important in understanding
the correlation patterns in CMB anisotropy when the integrated Sachs
Wolfe contribution is significant.

In the regime where $R_{\sc ls} < R_<$, there are no points within the
SLS that are topologically equivalent and the specific correlation
features discussed above are absent. However, quantitatively the CMB
correlations continue to have observable deviations from that expected
in a simply connected model until $R_{\sc ls}$ is substantially
smaller than $R_<$. Typically, the correlation pattern around any
point in the sky is anisotropic and distorted.

In simply connected models of the universe, the global homogeneity and
isotropy ensures that the CMB sky is statistically equivalent for all
observers. These global symmetries are generically absent in compact
(open or flat) universe models.  Consider two distinct observers.  If
one of them measures the correlation function between pairs $({\bf x},{\bf
x}^\prime$), the corresponding measurement for the other one will be
between pairs $(g {\bf x}, g{\bf x}^\prime)$ where $g$ is the element
of the isometry group of the universal cover, $g \in G^u$, which transports
the first observer to the position of the second one. (The motion $g$
always exists since the universal cover is homogeneous.)  For example if
${\bf x},{\bf x}^\prime$ belong to the SLS as seen by first observer,
then $g {\bf x}, g{\bf x}^\prime$ are on the SLS from the second
observer's point of view. Each pairwise correlation value $\xic$ is
determined by the set of distances ${\cal D({\bf x},{\bf x}^\prime)} =
\{d({\bf x}, \gamma_j{\bf x}^\prime)$, $\gamma_j \in \Gamma \}$. Moving
the observer translates this set to $\{d(g{\bf x}, \gamma_j g{\bf
x}\prime)\}$.  Since $G^u$ is the isometry group on the universal
cover, it conserves distances; in particular, $d(g{\bf x}, \gamma_j
g{\bf x}\prime) = d(g^{-1}g{\bf x}, g^{-1}\gamma_j g{\bf
x}\prime)$. Thus, under a general relocation of the observer by $g$,
the set of distances ${\cal D({\bf x},{\bf x}\prime)} $ is transformed
as $\{d({\bf x}, \gamma_j{\bf x}\prime)\} \rightarrow \{d({\bf x},
g^{-1}\gamma_j g{\bf x}\prime)\}$.

If $\Gamma$ is a normal/invariant subgroup of $G^u$,
($g^{-1}\gamma  g \in \Gamma, \forall \gamma\in\Gamma,g\in
G^u$), moving the observer simply reshuffles the terms in the set
${\cal D}$, and $\xic$=$\xi_{\Phi}^c(g{\bf x},g{\bf x}^\prime)$.
This is the case with simple tori, in which opposite sides are
identified by pure translation without rotation.  In the case of CH
spaces (as well as tori with twists), $\Gamma$ is not a normal
subgroup of $G^u$. Equivalent only are those observers mapped by a $g$
belonging to the isometry group $G$ of the compact space ${\cal M}$
itself,  not in the larger isometry group of 
its cover ${\cal M}^u$.  Each element of the isometry
group $G$ on ${\cal M}$ commutes with all the elements of $\Gamma$,
with the immediate result that the $C_{Tpp'}$ calculated will be the
same up to rotations of the sky if the observer is moved along the
orbits of $G$ in ${\cal M}$.

In $\S$~\ref{ssec_cmbcorr_varvar}, we discuss the correlation
pattern in the CMB that arises due to the global inhomogeneity of the
compact space.

\begin{figure}[h]
\plotone{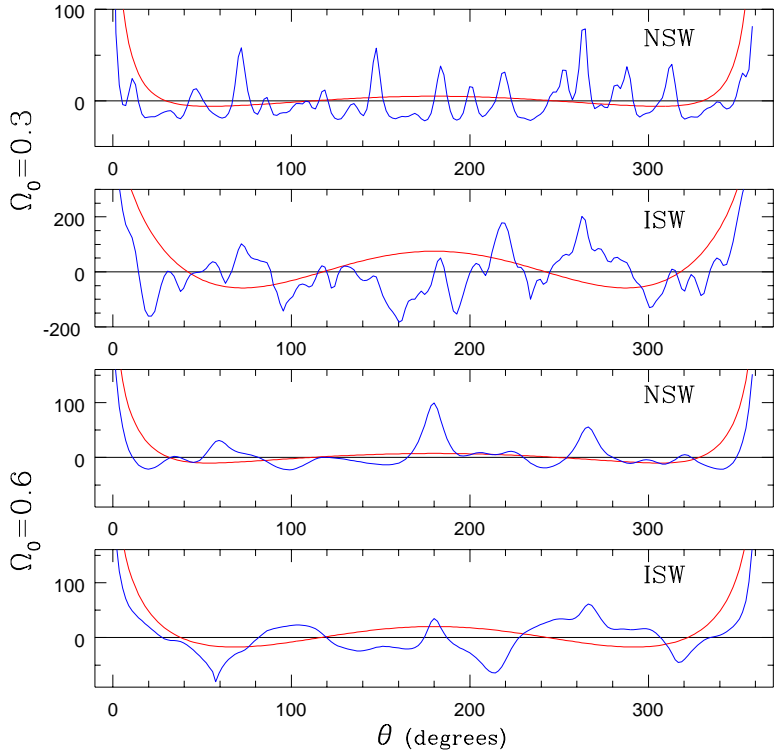}
\caption{ This shows the behavior of the correlation function of the
CMB temperature on a great circle in the sky in a CH model
(m004(-5,1)). The solid curves in the first and third panels show
$C(\theta)$ for the surface term in the Sachs-Wolfe effect (NSW) for
$\Omega_0=0.3$ and $0.6$, respectively.  They reflect the spatial
correlation along a circle on the sphere of last scattering (SLS). The
number of peaks in $C(\theta)$ matches the number of Dirichlet domains
that the circle intersects. The smooth curves show corresponding
results for the simply connected infinite $\hm$ models with
$\Omega_0=0.3$ and $0.6$. The second and fourth panels are analogous
to the first and third panels, except with the integrated Sachs-Wolfe (ISW)
effect included.  In this more physically correct case, the sharp
NSW peaks have been diluted by the ISW contributions. However, the ISW
effect induces new features; in particular, note the appearance of
strong negative correlations in panels two and four.}
\label{ctheta_line}
\end{figure}

\subsection{ Correlations due to the NSW Surface term alone}
\label{ssec_cmbcorr_nsw}

In Figure\,\ref{ctheta_line} the complex behavior of the correlation
function is illustrated with the example of the small CH model
(m004(-5,1)). The SLS encompasses $\approx 150$ domains for
$\Omega_0=0.3$ and $\approx 20$ domains for $\Omega_0=0.6$. It is
comparable to the size of one domain for $\Omega_0=0.9$. The angular
correlation along any arbitrary great circle in the sky in lower
$\Omega_0$ models shows distinct peaks as one encounters repeated
copies of the Dirichlet domain. The peaks are more pronounced when one
considers only the surface terms of eq.(\ref{dTSW}) in the CMB
anisotropy. Including the line of sight ISW contribution tends to
smearing the peaks, but also adds its own characteristic features as discussed
in $\S$~\ref{ssec_cmbcorr_isw}.

\begin{figure}[h]
\plottwo{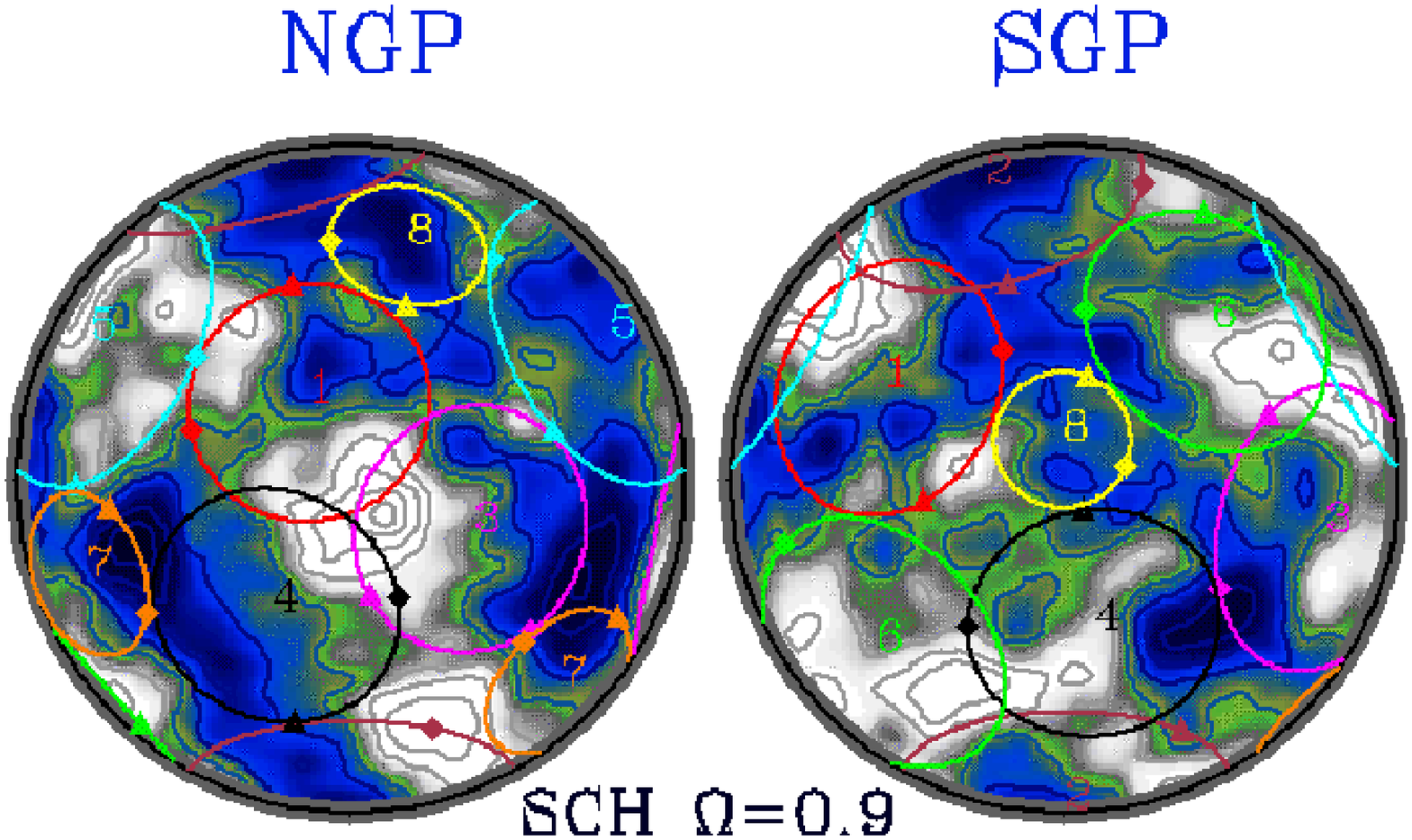}{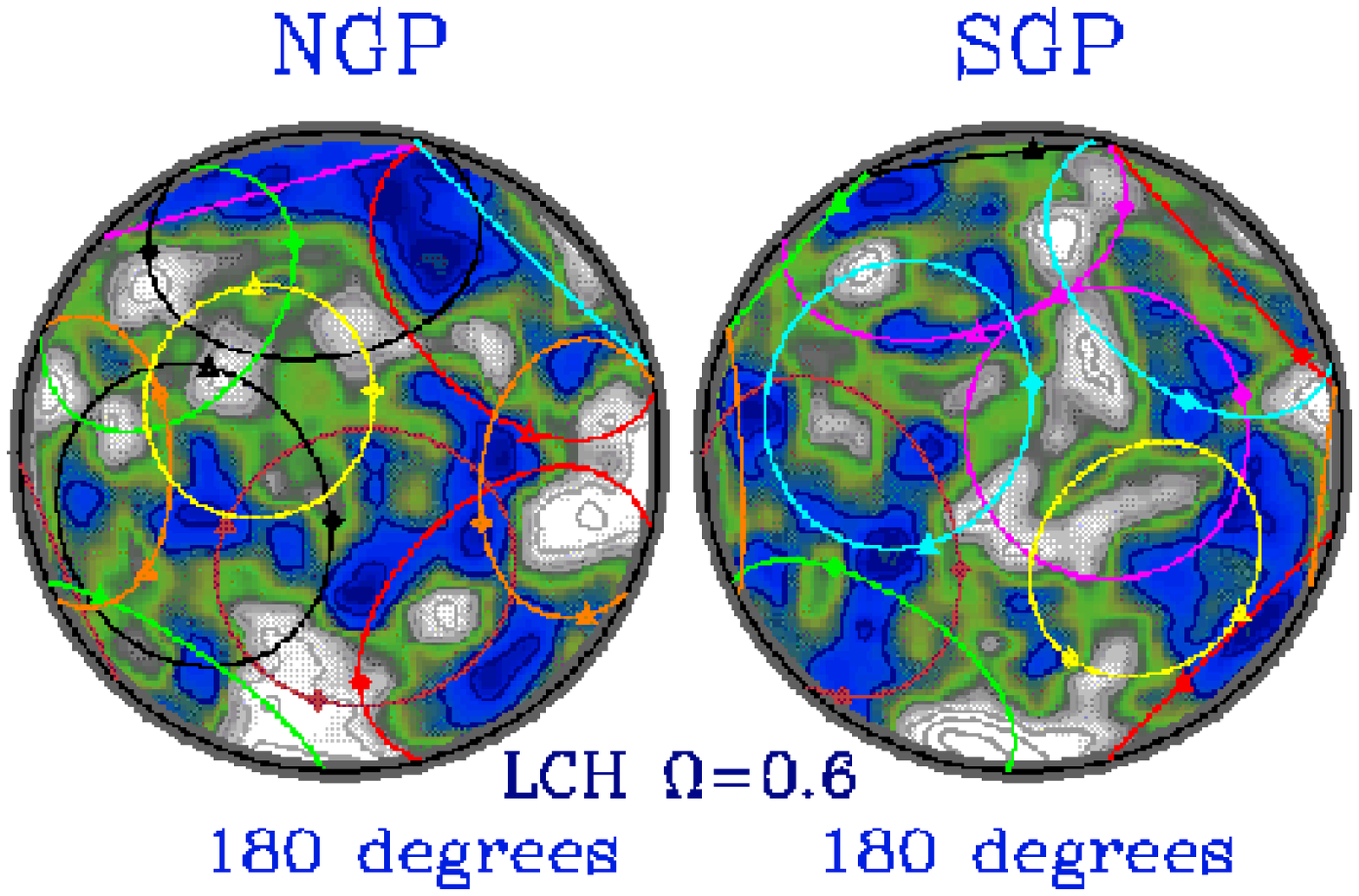}
\caption{ This shows two full-sky (noiseless) CMB anisotropy maps, plotted as
pairs of $180^\circ$ diameter hemispherical caps, one centered on the
South Galactic Pole (SGP) and one on the North (NGP). They are one of
an infinite number of possible random realizations based on the
computed pixel-pixel correlation matrix for the model in
question. Both surface and integrated Sachs-Wolfe effects have been
included. The power was normalized to best match the \cobedmr
data. The contours are linearly spaced in $30~\mu {\rm K}$ steps. In
contrast to Figs.~\ref{fig:v_isw} and \ref{fig:m_isw}, the maps are
not optimally filtered. The model labels $L(arge)CH$ and $S(mall)CH$
refer to the CH models v3543(2,3) and m004(-5,1), respectively. (The
model number associated with the topology corresponds to that of the
census of CH spaces from the Geometry center, Univ. of Minnesota;
$SCH$ is one of the smallest and $LCH$ is one of the largest spaces in
the census.) The value of $\Omega_0$ in each was chosen so that
$V_{\!\cal M}\sim V_{\sc sls}$. The matched pairs of circles expected
if the CMB anisotropy is dominated by the surface terms [28]
are superimposed on the map for each model. Each pair is labeled by
the same number centered on the circles. The relative phase is shown
by identified points marked by a diamond and a triangle on each circle
in a pair.  For clarity, we show only the eight largest pairs out of
$35$ for the $LCH_{\Omega_0=0.6}$ case. Even at \cobedmr resolution,
we find the cross-correlation between the temperature along matched
circles is very good in the $SCH_{\Omega_0=0.9}$ model.  The ISW
contribution is larger at $\Omega_0=0.6$, and the cross-correlation
coefficients are systematically smaller for $LCH_{\Omega_0=0.6}$
circle pairs.}
\label{fig:circles}
\end{figure}

Consider compact universe models which are small enough such that the
SLS does not completely fit inside one domain, $R_{\sc ls} > R_<$.
The CMB temperature is expected to be identical along pairs of circles
if temperature fluctuations are dominated by the surface terms at the
SLS~\cite{circles}. We identify these matched circles on the sky in
our models and check the extent of cross correlation seen at the
angular resolution of \cobedmr. Figure~\ref{fig:circles} shows the
matched pairs of circles in two (``small'' and ``large'') CH models
superimposed on random realizations of the theoretical sky generated
from the appropriate pixel-pixel correlation matrices. The models were
chosen to have a volume comparable to the volume $V_{\sc LS}$ within
the SLS. Even with the coarse pixelization of {\sc cobe--dmr}, we do
see fairly good cross-correlation in the CMB temperature along matched
circles in our realizations. Again, the pattern of correlated circles
is more pronounced when the Sachs-Wolfe surface term is dominant, as
happens in the small model ($SCH$) with $\Omega_0=0.9$.  We compute
the cross-correlation coefficient between the temperature fluctuations
along two matched circles $C_1$ and $C_2=\gamma C_1$, 
\begin{equation}
\rho_{12}=\langle\Delta T({\bf x}) \Delta T(\gamma {\bf x})\rangle/
\left[\langle\Delta T({\bf x})^2\rangle \langle\Delta T(\gamma{\bf x})^2\rangle
\right] ^{1/2}, ~ {\bf x} \in C_1, ~ \gamma {\bf x} \in C_2 \, .
\end{equation}
(In the case of a single realization  we replace the statistical average by the
integration over ${\bf x}$ along the circles.)
In the $SCH$ $\Omega_0=0.9$ case,  $\rho_{12}$ is in
the range $0.6-0.95$, whereas it is in
the range $0.2-0.6$ for the large model ($LCH$) with
$\Omega_0=0.6$.

The circles of identified pixels on the SLS are not the whole story.
Enhanced NSW cross-correlations, but at a lower level, also exist
between all pairs of points on the SLS which are projected close to
each other on the CH manifold. This can be seen as secondary maxima in
the examples in Figure\,\ref{ctheta_line}; these are absent in
standard cosmological models. Some features persist at a detectable
level even when the compact universe encompasses the SLS, \ie $R_{\sc
ls} \lta R_<$ , although circles are absent in this case (the effect
dies out for $R_{\sc ls} \ll R_<$).  When the relative ISW component
is significant, the geometrical patterns based on pointwise
identifications on the SLS are supplanted by more complex features
arising from identifications between photon geodesics, \eg the strong
negative correlations evident in Figure\,\ref{ctheta_line}.

\begin{figure}[tbh]
\plotone{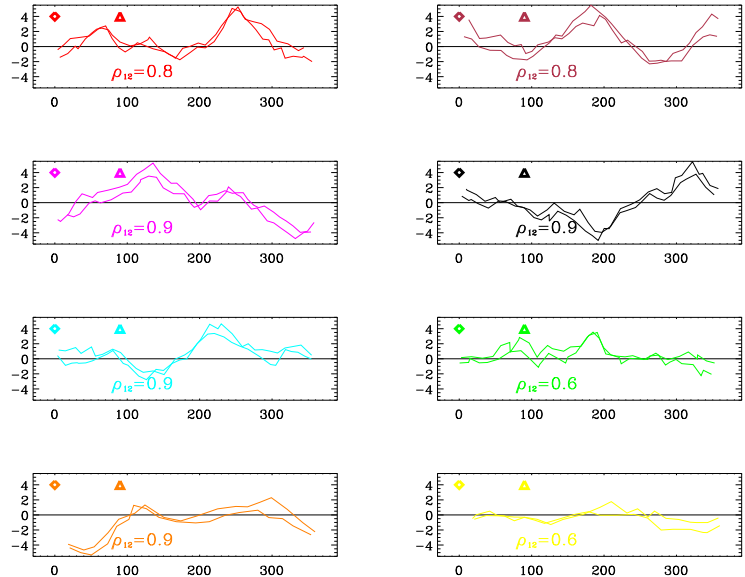}
\caption{ In the eight panels of the figure, the CMB temperature along
the $8$ matched pairs of circles of the SCH model shown in
Figure~\ref{fig:circles} (top panel) is plotted as a function of a polar
angle. The temperature shown is from the same realization of the CMB sky
that is used in Figure~\ref{fig:circles} and the diamond and triangles
are in correspondence. The CMB anisotropy includes both the surface
and integrated Sachs-Wolfe effects. At $\Omega_0=0.9$, the ISW
contribution is small, and thus the values of the correlation
coefficient are high.  }
\label{fig_corr_circ_m}
\end{figure}

\begin{figure}[tbh]
\plotone{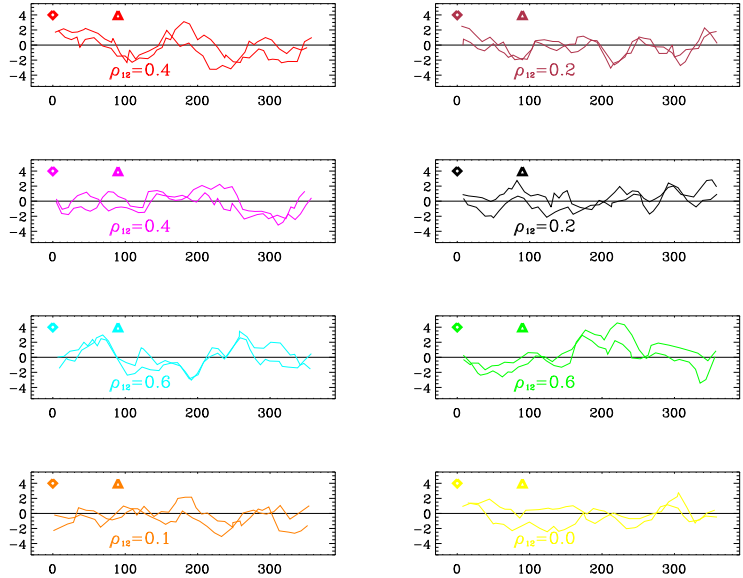}
\caption{ In the eight panels of the figure, the CMB temperature along
the largest $8$ matched pairs of circles of the LCH model
($\Omega_0=0.6$) shown in Figure~\ref{fig:circles} (bottom panel) is
plotted as a function of a polar angle. The temperature shown is from
the same realization of the CMB sky that is used in
Figure~\ref{fig:circles} and the diamond and triangles markers are in
correspondence. The CMB anisotropy includes both the surface and
integrated Sachs-Wolfe effects. At $\Omega_0=0.6$, the ISW contribution
is significant, resulting in small values of the correlation
coefficient. }
\label{fig_corr_circ_v}
\end{figure}

\subsection{ Correlations with the ISW effect included}
\label{ssec_cmbcorr_isw}

In a hyperbolic model, the temperature fluctuations depends on the
entire path of the photons from the sphere of last scattering to the
observer. The relative contribution of the ISW effect increases as the
value $\Omega_0$ decreases. As the ISW component increases, it tends
to wash out the patterns arising solely from the NSW (described in the
previous section). However, it also introduces additional features in
the CMB correlation. In Figures~\ref{fig_corr_circ_m} and
~\ref{fig_corr_circ_v}, the temperature fluctuations along matched
pairs of circles in a realization are plotted. Since $\Omega_0$ is
closer to unity in the small model ($SCH$), the temperature
fluctuations along the circles are more tightly correlated than in the
large model ($LCH$).

An example of ISW-induced features is the significant negative
correlations between widely separated directions in the sky, as is
clearly seen in the panels two and four of Figure~\ref{ctheta_line}.
In fact, the highly anti-correlated regions tend to lie at the centers
of the matched circles discussed in $\S$~\ref{ssec_cmbcorr_nsw}.
Figure ~\ref{fig_corrpixs} plots the anticorrelated pairs of pixels
separated by more that $\approx 10^\circ$ below a threshold in the
correlation coefficient of
$C_{pp\prime}/\sqrt{C_{pp}C_{p\prime p\prime}}=-0.3$ in the LCH model
with $\Omega_0=0.6$.  This should be contrasted with the highest
levels of anticorrelation of $\approx -0.04$ predicted in the
corresponding simply connected universe.

As discussed earlier, the lines of sight pointing to centers of
matched circles have segments of identical points and this can lead to
anticorrelation between the CMB anisotropy in those directions.  Here
we explain this with the help of Figure~\ref{fig:top}.  The segments
$AB^\prime$ and $A^\prime B$ of lines of sight $OA$ and $OB$ are
identical under $\Gamma$.  (Recall eqs.({\ref{cthetaSW}) and
({\ref{ctppSW}) for the Sachs-Wolfe contribution to the correlation
between temperature fluctuations in two directions in the sky.)  The
surface term depends on the correlation of the potential, $\Phi$,
between two points physically separated by the distance $r_1$.  The
interference term however contains correlation between identical
points, $AA^\prime$ and $BB^\prime$, and close ones.  The integral term
has correlations between close points but is down-weighted by the
extra factor of $\Delta F$.  Thus the interference term (which is
usually small in simply connected models) in this particular setting
can dominate the total CMB correlation and make it negative. Thus the
existence of strongly anticorrelated spots caused by the ISW term is a
signature of non-trivial topology.

\begin{figure}[h]
\plotone{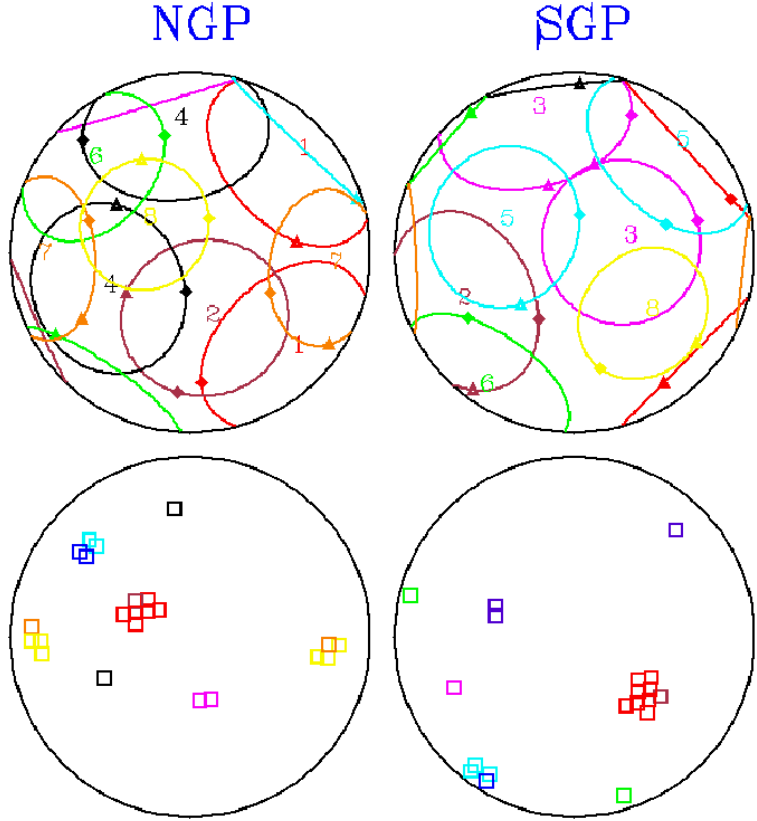}
\caption{ The negative correlation between pixels located at the
center of matched circles is demonstrated. The top panel shows the $8$
largest circle pairs for the LCH model on a sky map, as in
Figure~\ref{fig:circles}. The groups of pixels marked out in the lower
sky map are distant pairs with high anticorrelation,
$C_{pp\prime}/\sqrt{C_{pp}C_{p\prime p\prime}} < -0.3$.  It is visually
apparent that these regions lie at the centers of matched circles and
are labeled accordingly.  }

\label{fig_corrpixs}
\end{figure}

\subsection{Patterns Arising from Global Inhomogeneity}
\label{ssec_cmbcorr_varvar}

The global inhomogeneity of compact spaces implies that the variance
of the the gravitational potential is spatially dependent.  If these
fluctuations arose from quantum noise during inflation, $\Phi({\bf
x})$ would be an inhomogeneous Gaussian random field, with a pattern
of inhomogeneity determined by the topology of the compact space.

The CH models that we have considered also predict that the {\it rms}
temperature fluctuations in the sky vary with direction, as is shown
in Fig.~\ref{fig_varvar} and in Fig. 10 of Paper 1. In both LCH and
SCH models there are `loud' spots -- directions in the sky where the
variance is significantly larger. For the NSW contribution, these
arise when the SLS crosses loud regions in the Dirichlet domain where
the variance $\xico$ is large. These regions consist of points in the
Dirichlet domain where the length of the shortest geodesic is small
compared to the typical size of the shortest closed geodesic at other
points, i.e., the closest image of the point on the universal cover is
smaller than $\sim R_<$. The LCH model has a loud region around a
vertex of the Dirichlet domain shared by two pairs of identified faces
where the nearest image of the points close to this vertex are at a
small separation. We discussed this effect and results for the NSW
effect in a previous paper~\cite{paperA}.  Fig.~\ref{fig_varvar} shows
the extent to which the ISW term masks the signature expected from a
simple NSW consideration: the loud spot is dramatically modified, even
changing sign, when  ISW is included.

\begin{figure}[h]
\plotone{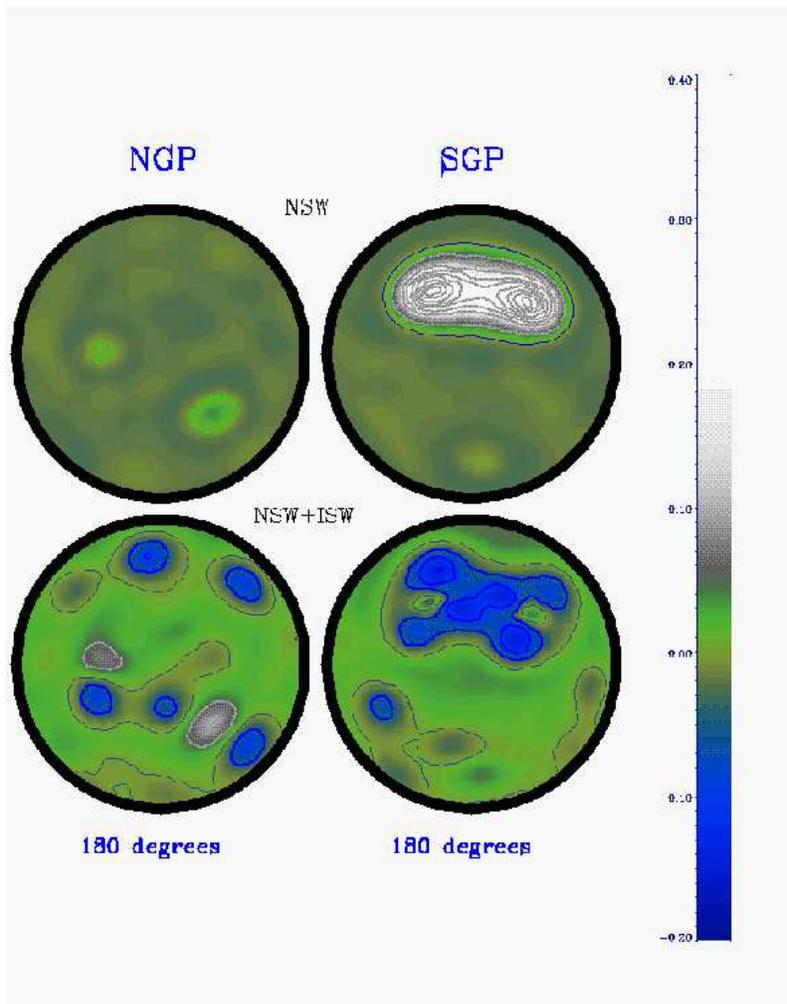}
\caption{ The figure shows two full-sky maps of the fluctuations in
the standard deviation of the predicted CMB temperature, plotted as pairs of
$180^\circ$ diameter hemispherical caps, one centered on the South
Galactic Pole (SGP) and one on the North (NGP).  The contours are
linearly spaced in units of 0.03 of the mean standard deviation. The first map
considers only the NSW contribution, where at the peak value standard deviation
is 45\% larger than its mean value . The second plot shows that
the significant ISW contribution at $\Omega_0=0.6$ radically alters
the pattern of fluctuations in the variance since the CMB temperature
now depends on the potential along the entire line of sight.  }
\label{fig_varvar}
\end{figure}

There are known examples of multiply connected spaces where the
pattern of inhomogeneity is a dominant observable effect. In
\cite{sok_star75} a non-compact but multiply connected horn topology
was constructed and the properties of the perturbations were
studied. It was demonstrated that there is a region in which the
variance of $\Phi$ is strongly suppressed and would correspond to a
dark spot in the distribution of galaxies. For the NSW contribution to
the CMB anisotropy, this same effect directly translates to
suppression of the variance of the temperature fluctuations in the
corresponding direction, creating a flat spot signature in the CMB
anisotropy, shown explicitly for the toroidal horn space
in~\cite{janna}.

\section{Bayesian analysis constraints from COBE-DMR }
\label{sec_prob}

In this section, the goal is to explicitly evaluate how likely the
various compact universe models are in light of the COBE data on
angular anisotropies. We first review the statistical distributions
for the maps derived from the COBE data, then show how we use our
techniques to confront the data. 

The raw data of a CMB experiment comes in the form of a time stream of
measurements $d_t$ at $N_t$ time-ordered points for each frequency
channel. The data are then binned into a $N_p$-pixel discretization of
the sky, through the relation $d_t = \sum_p P_{tp} \Delta_p + \eta_t$,
where the $N_t \times N_p$ pointing matrix $P_{tp}$ maps the observing
time to the angular position at that time and $\eta_t$ is the time
stream noise.  $\Delta_p$ is the true signal on the sky. With the
(reasonable and checkable) assumption that the noise is Gaussian with
covariance matrix $N=\langle\eta_t \eta_{t^\prime}\rangle$, one can
find the map $\bar\Delta$ which maximizes the conditional probability
${\cal P}(d|\Delta)$ and  the pixel-pixel noise covariance matrix 
about it, $C_N$.   
\begin{equation}
\bar\Delta = C_N P^\dagger N^{-1} d \, , ~~~~~ C_N=(P^\dagger N^{-1}
P)^{-1}\,,
\label{eq:bardelta}
\end{equation}
No larger moments
required, given  the Gaussian noise assumption. Thus, the probability
of the data given the true sky signal $\Delta$ is 
\begin{equation}
{\cal P}(\bar\Delta|\Delta) =
\frac{1}{(2 \pi)^{N_p/2} \|C_N\|^{1/2}} 
e^{-\frac{1}{2} (\Delta-\bar\Delta)^\dagger C_N^{-1} (\Delta-\bar\Delta)} \, , 
\label{eq:map}
\end{equation}
where $C_N=\langle(\Delta-\bar\Delta)(\Delta-\bar\Delta)^\dagger\rangle$. 
Provided the pixelization is fine-grained enough, the $\bar\Delta$ map
plus the $C_N$ contain all of the sky information 
present in the original data set $d_t$.

The COBE team\cite{dmr4} have given 6 maps at 3 frequencies, 31, 53,
and 90 GHz, each of 6144 pixels of size $(2.6^o)^2$, along with
information to construct $C_N$ from the number of observations made in
each pixel and the average noise in the radiometers over an observing
time. Four our analysis, we compressed the six \cobedmr maps
\cite{dmr4} into a (A+B)(31+53+90 GHz) weighted-sum map.  Galactic
emission near the plane of the Galaxy sufficiently contaminates the
primordial signal that a region $\pm 20^\circ$ from the Galactic plane
is removed, along with adjacent extra pixels in which contaminating
Galactic emission is known to be high, as advocated by the DMR team.
Although one can do analysis with the map's $(2.6^\circ)^2$ pixels,
this ``resolution 6'' pixelization of the quadrilateralized sphere is
oversampled relative to the \cobedmr beam size, and there is no
effective loss of information if we do further data compression by
using ``resolution 5'' pixels, $(5.2^\circ)^2$ \cite{bdmr294}.  The
celestial sphere is then represented by $N_p=1536$ pixels before the
Galactic cut, with $N_p=999$ pixels remaining after the cut is
made. Our $C_N$ is largely diagonal, but we include the off-diagonal
components centered on a $60^\circ$ pixel-pair angle separation, which
corresponds to the horn separation of the instrument.  We remove a
best-fit monopole and dipole from the cut-sky maps. Proper account is
taken of the monopole and dipole contributions, as well as possible
quadrupole contamination by Galaxy emission, by increasing the noise
in associated template patterns~\cite{BJ,BJK}. This corresponds to
having arbitrary monopole, dipole and quadrupole contamination
possible, and effectively ``shorts-out'' this contributions to $C_T$.

Now we need a probabilistic model for the signal. There may be several
physically different signals in the \cobedmr data - not only
primordial CMB anisotropy, but also Galactic emission and others.
However, except for the quadrupole contamination which we corrected
for, the contribution of signals other than CMB is small in the
weighted sum (over frequency channels) maps that we have used.  We have assumed
a Gaussian probability distribution for the primordial fluctuations of
$\Phi$. As we have seen $\Delta T/T $ is linearly related to $\Phi$,
which remains true even if all the effects leading to the CMB
anisotropy are included. Thus, $\Delta T/T $ is also statistically  a
Gaussian random field, fully described by the theoretical pixel-pixel
correlation matrix $C_T$ that we have focused on in our computations: 
\begin{equation}
{\cal P}(\Delta|C_T) = \frac{1}{(2 \pi)^{N_p/2} \|C_T\|^{1/2}}
e^{-\frac{1}{2} \Delta^\dagger C_T^{-1} \Delta} \, , ~~~~~~
C_T = C(\hat q_p,\hat q_{p^\prime})\,.
\end{equation}
The $C_T$ which enters here should have the COBE beam taken into
account.  We do this by calculating $C_T$ through eq.(\ref{ctppSW}). We used
a high $k$-cutoff in evaluating the $C^u$, choosing its value to
regulate the high frequency part of $C_T$, but ensured that it was much
higher than the corresponding scale associated with the COBE beam
size. The effect of the COBE beam was included by forming $B^\dagger
C_T B$, where $B_{pp^\prime}$ is the COBE beam between pixels $p$ and
$p^\prime$, related to the beam-shape $B_\ell$ by
\begin{equation}
B_{pp^\prime} = \sum_\ell {(2\ell +1 )\over 4\pi } B_\ell P_\ell (\hat{q}_p \cdot
\hat{q}_{p^\prime})  \, .
\end{equation}

This gives all the necessary ingredients for our analysis except for a
prior probability ${\cal P}(C_T)$ for the theory, which may encode
both our theoretical prejudices about the models and results from other
observational tests. For the moment we leave ${\cal P}(C_T)$
unspecified and concentrate on the likelihood function of a model
\begin{equation}
{\cal L}(C_T) \equiv {\cal P}(\bar\Delta| C_T) = \int d \Delta \,
{\cal P}(\bar\Delta | \Delta) \; {\cal P}(\Delta | C_T)\,.
\label{eq:likelihood}
\end{equation}
The integration is carried over all virtual realizations of the sky $\Delta$.
The result is 
\begin{equation}
{\cal L}(C_T)=
\frac{1}{(2 \pi)^{N_p/2} \|C_N+C_T\|^{1/2}}
e^{-\frac{1}{2} \bar \Delta^\dagger (C_N+C_T)^{-1} \bar\Delta }\,\,.
\end{equation}
The likelihood, defined by eq.(\ref{eq:likelihood}), is ultimately a
function of the parameters of the model, built into the
$C_T$. Maximization of ${\cal L}(C_T)$ in the parameter space is a
complex task, if, as is generally the case, the dependence on
parameters is nonlinear. However, it is straightforward to determine
relative likelihoods of any models with our precomputed $C_T$'s.

In the case of CH models the parameters are: the choice of manifold
${\cal M}$; density parameters, of the matter $\Omega_m$ and of vacuum
energy, $\Omega_\Lambda$ (most important is their combination
$\Omega_0$ which sets $R_{\sc ls}/d_c$, and, hence, the physical size
of the compact space); the manifold's orientation with respect to the
observed map - \ie the triplet of Euler angles ${\bf \alpha}$; the
position of the observer ${\bf x}_{obs}$ within the manifold; and the
parameters characterizing the initial spectrum of fluctuations, such
as the overall amplitude and the spectral tilt. If, as here, we fix
the initial spectral slope, only the amplitude $A$ remains free.  In
our studies, we have assumed a uniform prior probability for $A$ and
integrated the likelihood over it (\ie marginalized the parameter).
However, since ${\cal L}$ is always sharply peaked near the best-fit
value of $A$, the choice of the prior for $A$ is irrelevant.

Thus, ${\cal L}(C_T)= {\cal L}({\cal M},\Omega_m,\Omega_\Lambda,{\bf
\alpha},{\bf x}_{obs})$.  The logical way to proceed would be to do
many manifolds ${\cal M}$; for each manifold, many different
$\Omega_m$ and $\Omega_\Lambda$; for each $({\cal
M},\Omega_m,\Omega_\Lambda)$, many orientations ${\bf \alpha}$; etc.
For the exercise presented here we have chosen two model spaces, one
with a small volume, SCH, m004(-5,1), and one with a relatively large
onw, LCH, v3543(2,3), and considered `pure' open models
$\Omega_\Lambda=0$.  For each of these CH spaces we found the
likelihood for three values of $\Omega_m$ and twenty-four different
orientations.  We have chosen $\Omega_m$ values to straddle the line
$R_> \approx R_{\sc ls}$, since we have found this to be a rough
boundary between models which pass and which fail the \cobedmr
test. 

Twenty-four orientations correspond to the rotational symmetry
of a cube on which the \cobedmr pixelization is specified. This allows
us to avoid interpolating $C_T$ to new pixel positions for each
rotation and deal only with remapping of the correlation matrix
elements.  We have not varied the position of the observer. We have
chosen it to be at the ``local maximum of injectivity radius'', from
which position the space usually looks most symmetric (or `round'). We
expect that for this observer the model will be less restricted, than
for an observer at another place, who would see a more squashed and
anisotropic space. A caveat is if the most squashed direction is
partly hidden within the Galactic plane. 

In the torus model calculations, we chose many more manifold orientations to
sample the Euler angle space, since $C_T$ could be easily computed with a Fast
Fourier transform using the known eigenfunctions of the torus. In
general, we could continually refine our orientation angles to hone in
on the maximum likelihood value more precisely. We would obviously do
so if we felt that we were on the trail of a true model of the
Universe, but such a refinement is not essential for the points we
make here: that universes which are much smaller in volume than the
volume within the last scattering surface are strongly ruled out,
independent of orientation.

In Table I we present the results for the likelihood of the compact
models in our selection relative to the likelihood of the standard
non-compact open CDM model with the same $\Omega_m$.  The theory with
infinite volume is known to fit the \cobedmr data well and is
considered to give a good description of the data.
\cobedmr data alone does not discriminate well between the infinite
models with different values of $\Omega_m$ and $\Omega_\Lambda$, except
disfavoring very low $\Omega_m < 0.2 $, thus the choice of our
reference model is not critical.

\begin{table}[tbh]
\caption{The Log-likelihoods of the compact hyperbolic models relative
to the infinite models with the same $\Omega_m$ are listed. The
probabilities are calculated by confronting the models with the {\sc
cobe--dmr} data. The values quoted are likelihoods marginalized over
the amplitude of the initial power spectrum. The volume within the
sphere of last scattering (SLS) relative to the volume of the compact
models of the universe ($V_{\sc ls}/V_{\!\cal M}$) is listed.  The
three columns of logarithm of likelihood ratios ${\cal L}/{\cal L}_0$
correspond to the best, next best and the worst values that we have
obtained amongst $24$ different rotations of the compact space
relative to the sky. The number $\nu$ in brackets gives the
conventional, albeit crude, translation of the probabilities to a
Gaussian likelihood ${\cal L}/{\cal L}_0 \sim \exp[-\nu^2/2]$. Only
the last model for one specific orientation appears to be consistent
with the \cobedmr data.}
\begin{tabular}{cccccc}
\lbl{tab:logprob}
CH Topology &$\Omega_m$&$V_{\rm LS}/V_{\!\cal M}$& 
\multicolumn {3}{c}{Log of Likelihood Ratio (Gaussian approx.)}\\
$[V_{\!\cal M}/d_c^3,R_>,d_{\!\cal M}/d_c]$& & &\multicolumn {3}{c}{Orientation}\\
 &      &  &`best'&`second best'&`worst'\\
\tableline
 &0.3& 153.4 & -35.5 (8.4$\sigma$) & -35.7 (8.4$\sigma$)
& -57.9 (10.8$\sigma$)\\
{\bf m004(-5,1)} &0.6&19.3& -22.9 (6.8$\sigma$) & -23.3 (6.8$\sigma$)
& -49.4 ( 9.9$\sigma$)\\
$[0.98,0.75,0.86]$  &0.9&1.2 & -4.4  (3.0$\sigma$) & -8.5  (4.1$\sigma$)
& -37.4 ( 8.6$\sigma$)\\
\tableline
{\bf v3543(2,3)} & 0.6&2.9 & -3.6 (2.7$\sigma$) & -5.6 (3.3$\sigma$)
& -31.0 (7.9$\sigma$) \\
$[6.45,1.33,1.90]$& 0.8 &0.6&2.5 (2.2$\sigma$) & -0.8 (1.3$\sigma$)
& -12.6 (5.0$\sigma$) \\
\end{tabular}
\end{table}

The clear conclusion to be drawn from the Table is that when
$\Omega_m$ is too small, so $R_{>} < R_{\sc ls}$, the likelihoods are
tiny relative to the larger models. We believe this is a robust
conclusion, largely independent of the details of the manifold choice,
orientation or the assumed spectrum of the initial fluctuations.

What is also clear is that, near $R_{>} \approx R_{\sc ls}$, it can
happen that for certain manifolds and orientations, ${\cal L}$ is
higher than for the standard oCDM Universe. The interpretation is that
some of the highly correlated spots that are predicted, such as those
shown in Fig.~\ref{fig_corrpixs}, partly line up with the observed
spots in the \cobedmr map. Even though there are
realization-to-realization fluctuations, the random skies, derived
from an anisotropic model with correlated spots built into $C_T$, will
always be constrained to deliver pixel pairs reflecting these
correlations.  By contrast, in infinite isotropic models there are no
preferred spots and in a much smaller fraction of realizations
particular spot line-up will happen. Thus, this particular CH model at
the specfic orientation would be always preferred over its isotropic
infinite counterpart.

Allowing the manifold and its orientation to vary, we can get this
alignment from time to time, given so many parameters.  An obvious
question is how to assign the prior probability for orientation. This
should obviously be random, except that if we actually do live in CH
universe, there is a true orientation and we are not allowed to
marginalize over orientation nor manifold choice.  If high likelihood
is achieved for some manifold at a specific orientation, one could
argue that this model is a preferred explanation, at least for the
\cobedmr data.  What is then required to test this explanation?
Clearly, a strong test is to go to higher resolution. If the same
manifold and orientation remain preferred at higher resolutions, this
should spur cosmologists on to further checks of the CH hypothesis and
search for specific signatures of the compact space.  A powerful check
is to search for the correlated circles such as in
Figure~\ref{fig:circles}. A manifold-independent strategy with $13$
arcmin MAP data emphasized by Cornish \etal~\cite{circles} exploits
these correlated circles. A caveat is that there are other ``surface
terms'' involving the Doppler term which will spoil somewhat the
simplicity of this strategy.

Figures~\ref{fig:v_isw} and \ref{fig:m_isw} compare theoretical
realizations of the CMB anisotropy in the $LCH$ and $SCH$ models with
the \cobedmr data. They should be compared with the `DATA' map in
Fig.~\ref{fig:v_isw}, a Weiner-filtered picture of the CMB data.  The Wiener-filtered map
is the mean signal subject to the constraint of the observations for a
theory characterized by a given $C_T$: 
\begin{equation}
\langle\Delta_S|\bar\Delta,C_T\rangle = [C_T
(C_N+C_T)^{-1}]\bar\Delta \, .
\end{equation} 
The maps obtained with different choices for $C_T$ in the Wiener
filter often look quite similar {\it as long as the $C_T$ fits the
data reasonably well}. For $C_T$, we used that for a standard
$\Omega_m=1$ CDM model, which fits the data rather well. As far as the
visual appearance of the map is concerned, a $\Omega_m< 1$ oCDM model
would look very similar~\cite{bon_LH}. The differences are in line
with what we might expect: if there is a little more power on small
scales, then the map has slightly more contours at small angles.  When
one uses a model which is greatly disfavored by the data, the Wiener
map looks extremely different. For example, there is definitely a
pronounced large-scale signal in the DMR data which a low $\Omega_0$
compact hyperbolic model cannot reproduce.  It then tries to
interpret that large-scale signal as a chance (and highly unlikely)
superposition of noise, which is another expression of why the model
is so statistically disfavored.

What one should be noting looking at the maps is the shapes of the
patterns and not the specific locations of the patterns, since these
can change from realization to realization.  The full Bayesian
analysis takes into account all possible realizations.  The
incompatibility of models with small $V_{\cal M}/V_{\sc ls}$
(SCH-$\Omega_0=0.3,0.6$) is visually obvious: the best fit amplitudes
are high which is reflected in the steeper hot and cold
features. Although, the SCH-$\Omega_0=0.9$ and LCH-$\Omega_0=0.6$
models do not appear grossly inconsistent, it turns out that the
intrinsic anisotropic correlation pattern is at odds with the data
statistically (Table I).
\begin{figure}[tbh]
\plotone{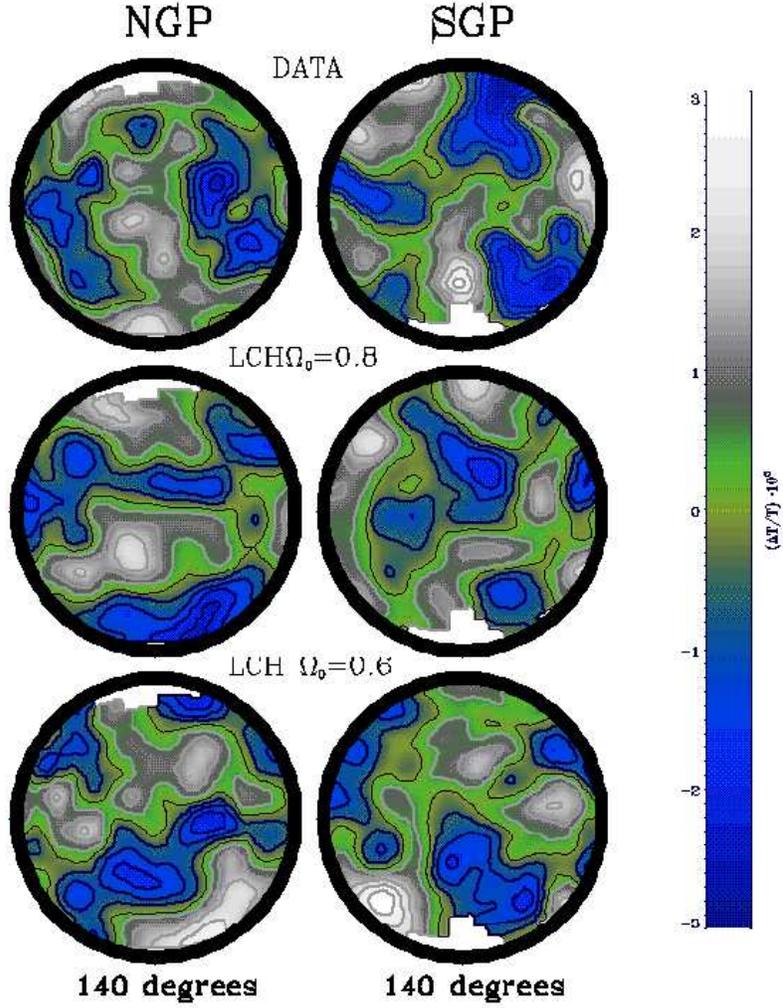}
\caption{ The figure consists of a column of three CMB sky-maps, each
showing a pair of $140^\circ$ diameter hemispherical caps, centered on
the South (SGP) and North (NGP) Galactic Poles, respectively. The top
map labeled DATA shows the \cobedmr 53+90+31 GHz A+B data after Wiener
filtering, assuming a best-fit standard CDM model, normalized to the
\cobedmr amplitude. The next two maps are of one random realization of
the CMB anisotropy in v3543(2,3) -- our choice of a $L(arge)CH$ model
example, for $\Omega_0=0.6~\&~0.8$ based on our theoretical
calculations of $C(\hat q,\hat q^\prime)$ convolved with the \cobedmr
beam.  Both surface and integrated (ISW) Sachs-Wolfe effects have been
included in $C(\hat q,\hat q^\prime)$. No noise was added. The
amplitude in each model was chosen to best-match the \cobedmr
data. The theoretical sky was Wiener-filtered using the \cobedmr
experimental noise to facilitate visual comparison with the data.  LCH
with $\Omega_0=0.8$ is compatible with the data with a suitable choice
of orientation, whereas with $\Omega_0=0.6$, it is ruled out (See
Table~\ref{tab:logprob}). For all the maps in Figs.~\ref{fig:v_isw}
and \ref{fig:m_isw}, the average, dipole and quadrupole determined for
this cut sky were removed. A $20^\circ$ Galactic
latitude cut was used, with extra pixel cuts to remove known regions of
Galactic emission proposed by the \cobedmr team, accounting for the
ragged edges. The contours are linearly spaced at $15~\mu {\rm K}$
steps.  The maps have been smoothed by a $1.66^\circ$ Gaussian
filter.}
\label{fig:v_isw} 
\end{figure}

\begin{figure}[tbh]
\plotone{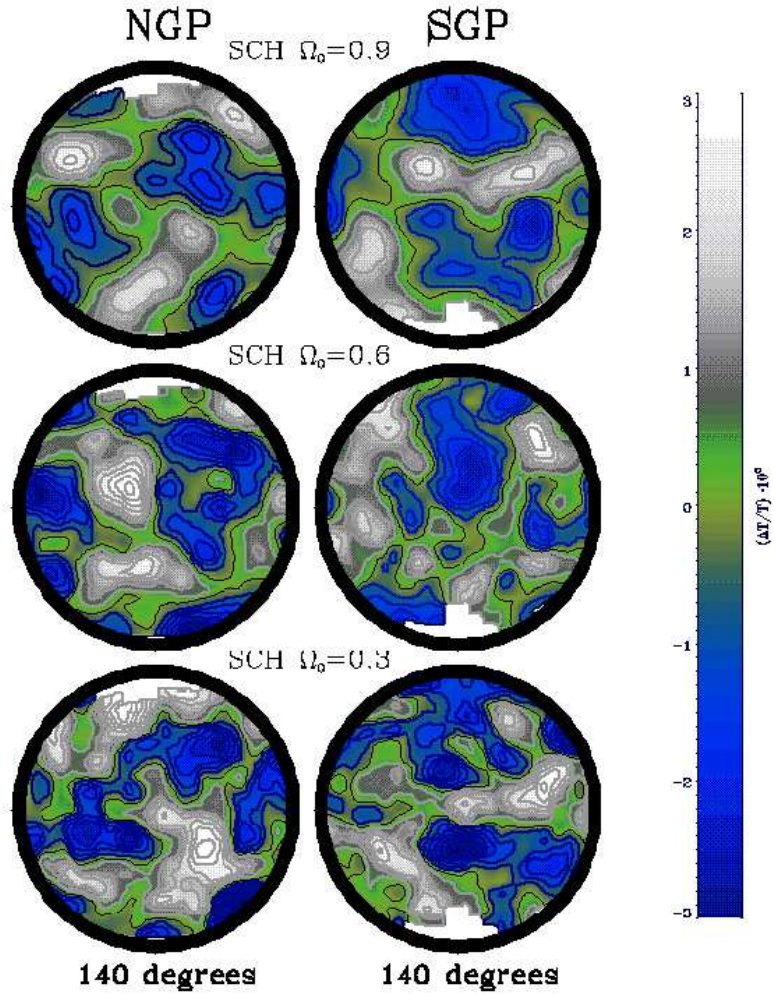}
\caption{ The three CMB sky-maps, each showing a pair of $140^\circ$
diameter hemispherical caps, centered on the South (SGP) and
North (NGP) Galactic Poles, are analogous to the lower two plots in
Fig.~\ref{fig:v_isw} but for the $S(mall)CH$ model m004(-5,1).  The
fact that $\Omega_0=0.3~\&~0.6$ models above are strongly ruled out by
the \cobedmr data is obvious visually. The $\Omega_0=0.9$ model 
is not obviously excluded on the visual basis, but is  indeed 
excluded on the basis of our Bayesian analysis.  (See
Table~\ref{tab:logprob}.)}
\label{fig:m_isw} 
\end{figure}

\section{Conclusion}

Although there are an infinite number of possible CH spaces, one can
extrapolate some general conclusions on CH universe models from our
limited exploration of \cobedmr constraints on the $LCH$ and $SCH$
examples.  We shall present the CMB constraints on a larger set of CH
spaces in \cite{us_prl}. The main CMB feature of small compact
universe models -- the presence of high correlations between many
well-separated pixel pairs -- is also their handicap. The \cobedmr
data does not, generally, favor interpretation as being a noisy random
realization derived from a small compact model.  Formally, the
likelihood ${\cal P}(\bar\Delta_{\sc cobe-dmr}|C_{Tpp^\prime})$ is
much smaller for such models than for a standard oCDM theory in which
$C_{Tpp^\prime}$ mostly just falls off with separation between the
pixels.

High correlations at large angles are numerous in a compact space with
$d_{\cal M} < R_{\sc ls}$ and we are confident in concluding that such
topologies are not viable models for our Universe in view of the \cobedmr
data.  Of course, the possibility of exceptions in the infinite list
of CH models remains, but the bulk of models must satisfy the, in our
opinion quite solid, limit $d_{\cal M} > \alpha R_{\sc ls}$ to
pass the CMB test.  Setting $\alpha=1$ would be a very conservative
choice; all our numerical simulations are
consistent with at least $\alpha=1.4$. At this limiting
value the CH models we tested are excluded at $3\sigma$ level.

Similar conclusions were reached by some of the authors (JRB, DP and
Igor Sokolov~\cite{us_torus}) for flat toroidal models. Comparison of
the full angular correlation (computed using the eigenfunction
expansion) with the \cobedmr data led to a much stronger limit on the
compactness of the universe than limits from other
methods~\cite{star_angel,tor_refs}. The main result of the analysis
was that $R_T/R_{\sc ls} > 1.3$ at $95\%~CL$ for the equal-sided
$3$-torus (with the periodicity length $2R_T$, the diameter of the
torus is $d_T=3^{1/2} R_T$, thus $\alpha=2.25$ ).  For $3$-tori with
only one short dimension (or for the non-compact $1$-torus), the
constraint on the most compact dimension is not quite as strong
because the features can be hidden in the ``zone of avoidance''
associated with the Galactic cut.

Statistical properties of fluctuations in CH manifolds are anisotropic
and inhomogeneous. CMB predictions of course depend not only on the
topology of the space, but also on the position of the observer and
the orientation of the Dirichlet domain with respect to \cobedmr sky.
This lack of high symmetry is reflected, in particular, in $R_>$
and $R_<$, which are not invariant under the choice of the space
basepoint (\ie the observer position) and which quantify how the sky
looks for a given observer. Our limit on CH models uses the invariant
linear measure $d_{\cal M}$, but roughly corresponds to the condition
$R_> > R_{\sc ls}$ when $R_>$ is the outradius for the observer at the
basepoint which maximizes the `injectivity radius' of the
space~\cite{Minn}. This means that when the Dirichlet domain just fits
into the last-scattering sphere, the correlation matrix is already too
distorted to satisfy the data.  Moving the observer to another point,
will generally increase $R_>$, but will also squash the domain in some
other directions. In the more anisotropic view of the CH space
presented to such an observer, we expect to predict less favorable CMB
skies. Thus, moving the observer away from the basepoint which
maximizes the `injectivity radius' may not relax the constraint. The
absence of good data close to the galactic plane, (\ie the galactic
cut in the data) may help some models at specific orientations, but
not to the extent that it does for the $1$-torus space, which has an
exact planar symmetry. Extensive analysis of the changes induced by
varying observers is left to future work.

As we emphasized in \cite{us_texas,us_cwru} and here, the constraints
arise predominantly from predicted pattern mismatches in our models
compared with the COBE tapestry. This is entirely encoded in
$C_{Tpp^\prime}$, which can also be expressed in terms of a $Y_{\ell
m}$ basis. However, $C_{T\ell m, \ell^\prime m^\prime}$ is generally
quite complex and reducing consideration to the isotropized
${\cal C}_\ell$ loses a substantial amount of information, since it
involves a $\delta_{\ell \ell^\prime}$ projection, followed by a trace
over $m$. More importantly, as we show in detail in
Appendix~\ref{sec_noCl}, ${\cal C}_\ell$ has substantially increased
error bars from ``cosmic variance'', \ie in the expected theoretical
fluctuations about the mean, so we can draw only extremely weak
conclusions about the model. This is evident in the
error bars on the angular power spectrum shown for a set of the CH
models in Figure~\ref{fig:clCH}.  Thus, although the ${\cal C}_\ell$ for a
compact model may fit the data reasonably well, and it sometimes does
so even better than the corresponding infinite model with the same
$\Omega_0$, statistically this may be a rather meaningless
observation, and, if one is not careful, even misleading. Some
authors~\cite{corn_sper99,inoue99b} have argued that because 
the mean ${\cal C}_\ell$ shape may look better  visually, this is
evidence that the models are preferred. 

To make the point quantitatively that conventional use of ${\cal
C}_\ell$ can lead to very wrong conclusions,
Table~\ref{tab:isologprob} compares the likelihood ratios for a few
models obtained using just ${\cal C}_\ell$ information, treated as if
they were statistically isotropic Gaussian models with the the
isotropized power spectrum, with what was obtained in
Table~\ref{tab:logprob} when the full pattern-recognition statistical
treatment was made. Details on the construction of the Table are given
in Appendix~\ref{sec_noCl}. In all cases shown, the CH ${\cal C}_\ell$
is preferred over the ${\cal C}_\ell$ of the corresponding infinite
model, but it is grossly misleading because of the enhanced error
bars, and the huge amount of relevant information left out. All models
are {\it strongly} ruled out, save one. And that manifold, with its
specific orientation relative to the sky, is preferred even more than
the statistically-isotropized one.

\begin{figure}
\plotone{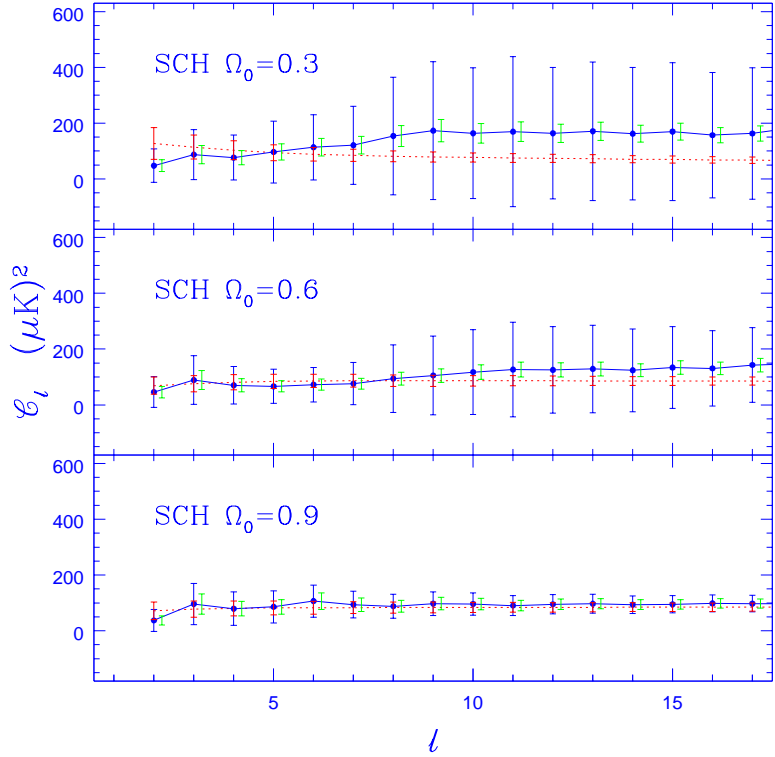}
\caption{The three panels show angular power
spectra for the SCH model at $\Omega_0=0.3$, $0.6$ and $0.9$,
respectively. The larger error bars are the actual cosmic variance
computed using the full correlation matrix information. The smaller
error bars drawn slightly displaced to the right are the cosmic
variance one would assign if one naively assumed that the ${\cal
C}_\ell$ contained all the information. The excess variance reflects
the incompleteness of the ${\cal C}_\ell$ information due to the
inherent statistical anisotropy. The dotted (red) curve shows 
${\cal C}_\ell$ for the simply connected `open' universe at the same
value of $\Omega_0$. The error bars show the cosmic variance. The
${\cal C}_\ell$ is normalized to give the same {\it r.m.s.} power as
the corresponding SCH model for the \cobedmr beam. 
The full \cobedmr beam, with rough Gaussian scale $\sigma_{\rm
beam}^{-1} \sim 17.5$, 
has been factored out in making the plot.}
\label{fig:clCH}
\end{figure}

\begin{table}[h]
\caption{The Log-likelihoods of the compact hyperbolic models relative
to the infinite models with the same $\Omega_m$ are listed. The
probabilities are calculated by confronting the models with the
\cobedmr data solely in terms of ${\cal C}_\ell$. For easy reference
the corresponding relative likelihood for the ``best'' orientation is
listed from Table~\ref{tab:logprob}.}
\begin{tabular}{cccccc}
\lbl{tab:isologprob} 
CH Topology &\multicolumn {3}{c}{{\bf SCH: m004(-5,1)}}& \multicolumn
{2}{c}{{\bf LCH: v3543(2,3)}}\\
$\Omega_m$ &0.3 & 0.6 & 0.9 & 0.6 & 0.8\\
\tableline 
Log. Likelihood Ratio &0.62(1.1$\sigma$)& 0.43(0.93$\sigma$) &0.48(0.98$\sigma$) &0.61(1.1$\sigma$)&0.82(1.3$\sigma$) \\
Using only ${\cal C}_\ell$&&&&\\
&&&&\\
Log. Likelihood Ratio &-35.5(8.4$\sigma$)& -22.9 (6.8$\sigma$) &-4.4(3.0$\sigma$) &-3.6(2.7$\sigma$)&2.5(2.2$\sigma$) \\
Using full $C_{Tpp^\prime}$ (``Best'' orientation)&&&&\\
\end{tabular}
\end{table}

Most astrophysical observations point to a matter density $\Omega_m
\lta 0.4$~\cite{opencase}. For the matter-dominated
$\Omega_0=\Omega_m$ open cosmology, when we combine this with the
$d_{\cal M} > \alpha R_{\sc sls}$ \cobedmr constraint that we suggest,
\ie $\Omega_m > 1 - {\rm tanh}^2 \left(d_{\cal M}/2 \alpha d_c\right)
$, we would be able to rule out topologically small CH models with
$d_{\cal M}/d_c \lta 2.8$ (adopting the conservative $\alpha=1.4$). 

Recent SNIa results~\cite{SNres}, the emerging location of a peak in
the CMB power spectrum at $\ell \sim 200$, and the combination of CMB
with large scale structure data~\cite{BJroysoc} all point to a
significant $\Omega_\Lambda$ term, with less room for a small
$\Omega_{0} = \Omega_{m}+ \Omega_{\Lambda}$, even though $\Omega_m$
may be small. Ironically, although this makes open models less
attractive, having a non-zero $\Omega_{\Lambda}$ which combines with
$\Omega_m$ to give $\Omega_{0}$ near unity actually returns
topologically small CH spaces back to life, since it alleviates the
constraint on the size $d_{\cal M}$.  Qualitatively, CH models with
the same value of $R_{\sc ls}/d_c$ will have similar constraints,
whether there is $\Omega_\Lambda$ or not. Some quantitative
differences will of course arise because of the diminished
contribution from the ISW effect in the $\Omega_m + \Omega_\Lambda$
model compared to the corresponding $\Omega_\Lambda=0 $ one. 
  
In Figure~\ref{fig:limits} we show the $\Omega_m - \Omega_\Lambda$
parameter space, where the lines of constant $R_{\sc ls}/d_c$ provide
rough limits on the viability of the CH models, depending on its size.
This plot illustrates that allowing for nonzero $\Omega_\Lambda$
relaxes the limits on the allowed CH topology. This is a welcome
conclusion, since one could argue that (topologically) small CH spaces
are less complex and may be more probable for quantum processes in the
early Universe to have created them, making them a more natural choice
among other CH models~\cite{gib98}.

\begin{figure}
\plotone{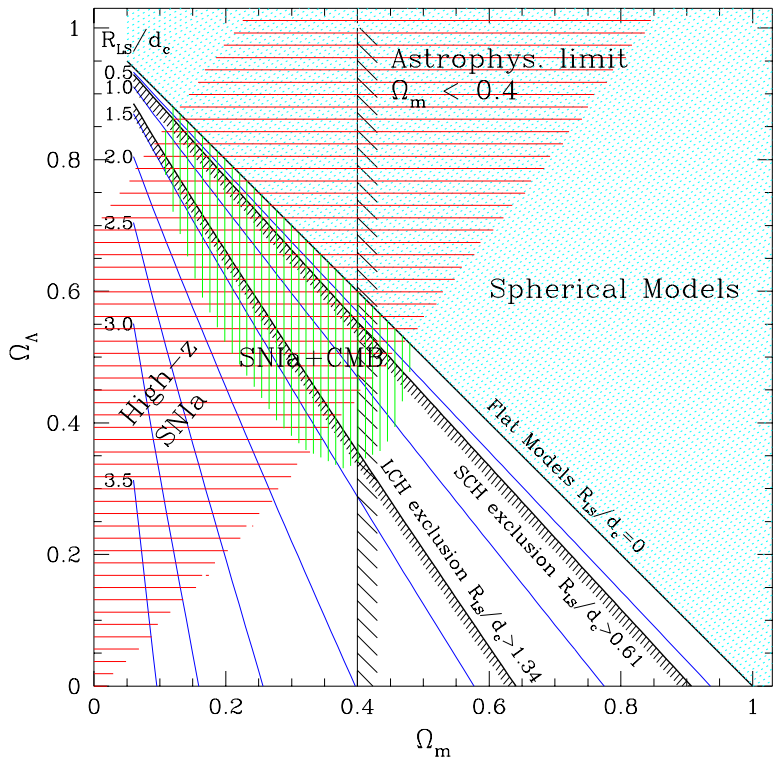}
\caption{In the $\Omega_m - \Omega_\Lambda$ parameter plane, the
horizontally shaded area shows the 95\% CL (confidence limit) region
coming from the analysis of high redshift supernovae SNIa [10].
For reference, the vertically shaded region shows 95\%
CL restrictions arising in the standard infinite universes when SNIa
and CMB data are combined. Since intermediate-scale CMB data dominates
the CMB constraint on the first peak position and hence on $\Omega_0$,
where the influence of the topology would not be so important, the
limits on $\Omega_m, \Omega_\Lambda$ in CH spaces should be similar to
those shown. We suggest that lines of constant $R_{\sc ls}/d_c$ would
provide effective guides to the viability of the compact models.  The
two lines with enhanced weight correspond to the line $R_{\sc
ls}/d_c=d_{\cal M}/1.4$ in our SCH and LCH example spaces.  The
allowed region is to the right of the $R_{\sc ls}/d_c=d_{\cal M}/1.4$
lines. The vertical line at $\Omega_m=0.4$ cuts out the high matter
density part of the plane which is disfavored by observations.  }
\label{fig:limits}
\end{figure}

Although our results strongly indicate that manifolds with small $R_>
< R_{\sc ls}$ are unlikely to survive confrontation with the \cobedmr
data, we emphasize that, in the $R_> > R_{\sc ls} > R_< $ regime, there
is some room both to have interesting specific CH correlation
patterns and still be consistent with the \cobedmr data.  When $R_{<}$
is large compared to $R_{\sc ls}$, the results will quickly converge
towards the usual infinite hyperbolic manifold results. The
intermediate terrain still encompasses ample scope for interesting
topological signatures to be discovered within the CMB. Although our
methods are quite general, testing all manifolds in the SnapPea census
this way is rather daunting, and there are countably infinite
manifolds not yet prescribed. What may be promising for discovery are
specialized statistical indicators, which are less powerful discriminators
than the full Bayesian approach we have used here, but not as manifold
sensitive; {\it e.g.}, the statistical techniques which exploit the
high degree of correlation along circle pairs that \cite{circles}
have emphasized, and Fig.~\ref{fig:circles} reveals. Maps like we have
constructed will be necessary to test the statistical significance of
such methods. We also note that dramatically increasing the resolution
beyond that of \cobedmr, to e.g. MAP resolution,
is quite feasible with current computing power using our techniques.

\section{Acknowledgements}
We have made extensive use of the SnapPea package and related material
on compact hyperbolic spaces, which is available at the public website
of the Geometry center at the University of Minnesota. TS acknowledges
support during the final stages of this work from NSF grant
EPS-9550487 with matching support from the state of Kansas. In the
course of this project we have had enjoyable interactions with Janna
Levin, Neil Cornish, Dave Spergel, Igor Sokolov, Glen Starkman and
Jeff Weeks.

\appendix

\section{Incorporating small angle CMB anisotropy }
\lbl{sec_cthetafull}

In this section, we discuss the calculation using the method of images
of the primary CMB anisotropy at smaller angular scales where
sources other than the Sachs-Wolfe effect make the dominant
contribution. The most important effects are the Doppler shift due to
scattering of photons on free electrons during recombination, and a term
describing the compression and rarefaction of photons.

We shall consider only the scalar mode of perturbations.  As in the main
text of this paper, we choose to work in the longitudinal gauge in
which the metric perturbations are described by two scalar potentials
$\Phi$ and $\Psi$

\begin{equation}
ds^2=a^2(\tau)\left[\left(1+2\Phi\right) d\tau^2 - (1-2 \Psi)
g^{(3)}_{ik} dx^{i}dx^{k} \right]\,.
\end{equation}
This gauge is particularly suitable for the analysis of multiply
connected universes, since the perturbed 3-space is explicitly
conformal to the background 3-space, leaving the topological
identification of points in the background coordinates $x^i$ exact.
Also for the perturbed quantities in the longitudinal gauge one does
not need to distinguish between gauge-specific and gauge-invariant
definitions.

In first-order perturbation theory, the photons observed at the
origin from the direction $\hat q$ at the (present) moment $\tau_0$
have propagated along radial null-geodesics. The position $(\tau,
x^i)$ on the geodesics is parametrized by $(\tau_0-\chi, \chi\hat q)$ 
in terms of the affine parameter $\chi$ with $\chi=0$ at the
origin. $\tau_0$ coincides with the present radius of the FRW horizon,
$\chi_H$.

The transport of the temperature fluctuation $\Delta_T
(\tau,x^k,\hat q )$ along the photon path is given by the
Boltzmann equation
\begin{equation}
-{\cal D}_\chi \Delta_T + (\partial_\chi\zeta)\,\Delta_T =
\left(\partial_\tau\Psi + \hat q^i \partial_i \Phi \right) +
(\partial_\chi\zeta)\, \left(\varepsilon/4 + \hat q^i \partial_i
{\psi_v}\right), ~~~~ {\cal D}_\chi = \hat q^i \partial_i
- \partial_\tau \,,
\label{boltzeq}
\end{equation}
where we have used the following notations~\footnote{Other frequently
used notations are $\nu=\Phi$, $\varphi=\Psi$,
$\delta_{\gamma}=\varepsilon$, as in \cite{bon_LH}}: ${\cal D}_\chi$
is the total derivative along the photon path and $\partial$ denotes
the partial derivatives.  Along the path the direction of the photon
momentum $\hat p$ is opposite to the direction of $\hat q$.  The
energy density fluctuations are given by local angle averaging over
momentum directions $\varepsilon(\tau,x^i)= \int d\Omega_{\hat p}
\Delta_T (\tau,x^i,\hat p)/4\pi$.  The function $\zeta(\chi) =
\int_0^{\chi} d\chi \sigma_T n_e $ is the optical depth due to
Thompson scattering of photons on free electrons with a number
density $n_e$. The velocity is described by the velocity potential
$\psi_v$. We have also omitted subdominant terms related to the
effects of polarization and the angular anisotropy in the
scattering~\cite{bon_LH}.

The resulting temperature fluctuation measured by the observer in the
direction $\hat q$ is obtained by integrating eq.~(\ref{boltzeq})
along the photon path
\begin{equation}
\Delta_T (\tau_0,\hat q) = \int_0^{\tau_0} {\mathrm
D}\chi\, \left[ \left(\partial_\tau\Psi +
\hat q^i\partial_i\Phi \right)\,e^{-\zeta(\chi)}
- \left(\epsilon/4 + \hat q^i\partial_i\psi_v\right)
\partial_\chi e^{-\zeta(\chi)} \right]\,.
\label{eq:boltzsol}
\end{equation}

Assuming the standard recombination history, the approximation of
`instant recombination' is quite accurate for the CMB anisotropy at
large and intermediate angular scales. It assumes an instantaneous
transition from the phase at $\tau<\tauls $, where the photons were
tightly coupled with the electrons, to the phase at $\tau >\tauls$
where they propagate freely after last scattering at $\tauls$.  In
eq.~(\ref{eq:boltzsol}) this formally corresponds to the limit for the
visibility function $ e^{-\zeta(\chi)} \to
\Theta(\tau_0-\tauls-\chi)$, where $\Theta(x)$ is the Heaviside step
function. Correspondingly, the differential visibility function $
-\partial_\chi e^{-\zeta(\chi)} \to \delta (\tau_0-\tauls-\chi)$,
where $\delta(x)$ is the delta function.  Omitting the monopole terms,
one obtains the well known expression for temperature fluctuations in
this limit:
\begin{equation}
\Delta_T (\tau_0,\hat q) = \left(\epsilon/4 + \Phi +
\hat q^i \partial_i\psi_v\right)
\left|_{\stackrel{{\scriptstyle\tau=\tauls}}{\chi=\tau_0-\tauls}}\right.
 +
\int_0^{\tau_0-\tauls} {\cal D}\chi\, \partial_\tau\left(\Psi +
\Phi\right) .
\label{eq:instant} 
\end{equation}
The above expression for the CMB temperature fluctuation includes the
Doppler term $\hat q^i \partial_i\psi_v$ in addition to the
Sachs-Wolfe contribution given in eq.~(\ref{dTSW}). To obtain the
Sachs-Wolfe contribution in the form of eq.~(\ref{dTSW}), the relation
$\Phi=\Psi$ valid in the (fairly good) hydrodynamic approximation to
the matter content in the universe is used.  Furthermore, the well
known solution to the evolution equations for the perturbations imply
$\varepsilon(\tauls)\approx -8\Phi(\tauls)/3$ in the Sachs-Wolfe
regime.

We have presented these well known results to stress that, in full
generality, the CMB anisotropy can be expressed as a line of sight
integral over a source function given purely in the real space. Thus,
in a multiply-connected Universe the method of images can be directly
applied to full CMB calculations in a similar manner as we have 
used it to calculate Sachs-Wolfe contribution.

The new element that eq. (\ref{eq:boltzsol}) has beyond the
Sachs-Wolfe expression of eq.~(\ref{dTSW}) is the
vectorial terms in the source function. For the scalar perturbations
all of them are spatial derivatives of the potentials along the line
of sight and can be reduced to time derivatives through integration by
parts. Indeed, $ \hat q^i\partial_i\Phi $ is readily
converted into a $ \partial_\tau \Phi $ addition to the integral plus
the surface term $ \Phi [\tauls,(\tau_0-\tauls)\hat q] $, as was
implicitly done both in eq. (\ref{eq:instant}) and in eq.~(\ref{dTSW}) in
the main text. The Doppler term can also be dealt with the same way,
however there is no obvious advantage of doing that. The reason is the
$\delta$-function like, rather than step-function like, nature of
the differential visibility multiplier in the Doppler term. Integration by
parts gives an integral term with $\partial^2_\chi e^{-\zeta(\chi)} $,
which quickly changes sign; and having to integrate it numerically may not
indeed be a simplification over numerical differentiation.

As is clear in the simplifying instant recombination approximation,
{\em the Doppler term depends on the line-of-sight projection of a vector
at last-scattering surface and not a scalar}.
As for a scalar field, a vector field on a multiply
connected universe ${\cal M}^u/\Gamma$, can be expressed as a sum over
images of the vector field on the universal cover, ${\cal
M}^u$. However, in contrast to scalar fields, the discrete group
$\Gamma$ acts both on the spatial position ${\bf x}$ and the vector at
that point, \ie the vector at the image point $\gamma[{\bf x}]$ is
rotated relative to the vector at ${\bf x}$; and its projection on the
line-of-sight is not preserved.  Thus, at scales where it
is dominant, the presence of the Doppler term at the last-scattering
surface will tend to destroy the nice circle correlations that appear
for the NSW case (see $\S$~\ref{ssec_cmbcorr_nsw}).

However, the presence of vector terms is not an obstacle for the
numerical implementation of the method of images. Indeed, it is quite
straightforward to implement the method of images to calculate the CMB
correlation function for eq.~(\ref{eq:instant}) or
eq.~(\ref{eq:boltzsol}).  For a completely general source term, the
correlation function for the CMB anisotropy is the double integral
\begin{equation}
C(\hat q,\hat q^\prime) = \int_{0}^\chiH d\chix{1}~\int_{0}^\chiH
d\chix{2}~\langle {\cal S} (\hat q \chix{1}){\cal S} (\hat q^\prime
\chix{2}) \rangle\,.  
\lbl{ctheta_full}
\end{equation}
In a compact (more generally, multiply connected) FRW universe, the
method of images can be invoked to compute the source correlation
function as a regularized sum over images over the source
correlation function on the universal cover:
\begin{equation}
\langle {\cal S} (\hat q \chix{1}){\cal S} (\hat q^\prime \chix{2})
\rangle^c = \widetilde{\sum_{\gamma\in\Gamma}} \langle {\cal S} (\hat q
\chix{1}) \gamma[{\cal S} (\gamma[\hat q^\prime \chix{2}])] \rangle^u
\label{moi_source}
\end{equation}
where the superscripts $c$ and $u$ refer to the quantity in the
multiply connected space and its universal cover, respectively.  Tilde
refers to the possible need for regularization.
$\Gamma$ is the discrete subgroup of motions which defines the
multiply connected space and $\gamma[{\bf x}]$ is the spatial point on
the universal cover obtained by the action of the motion
$\gamma\in\Gamma$ on the point ${\bf x}$. The important point to note
is that one needs to implement the action of the motion $\gamma$ on
the source function unless all the terms in the source function are
scalar quantities (as was the case for the Sachs-Wolfe effect that we
considered in this paper) when the action is trivial.

In summary, the CMB anisotropy correlation in a multiply connected
universe can be computed in full generality using the method of images
on the correlation function for CMB anisotropy source terms. However,
it is important to identify the scalar, vector and tensor parts of the
source functions and the action of the discrete motion has to be
applied to non-scalar components of the source function.

\section{Inadequacy of ${\cal C}_\ell$ comparison for compact universes}
\label{sec_noCl}

 In this appendix, we use the examples of the CH spaces we have
studied here to discuss the limitations of a comparison of CMB
anisotropy predictions to data solely in terms of the angular power
spectrum for compact universe models.

The (isotropized) angular power spectrum, ${\cal C}_\ell$, widely used
to summarize the CMB anisotropy predictions of a theoretical model, is
defined by
\begin{equation}
{\cal C}_\ell \equiv \frac{\ell(\ell+1)}{2\pi (2\ell +1)}~
\sum_{m=-\ell}^{\ell} \langle \hat a_{\ell m} \hat a^*_{\ell
m}\rangle,\quad {\rm where}\quad \hat a_{\ell m} = \int d\Omega_{\hat
q}\, \widehat\dT(\hat q)\, Y_{\ell m} (\hat q)\,.
\label{def_Cl}
\end{equation} 
The $\langle\,\rangle$ denotes an ensemble average of the random
variable enclosed. In full generality, the expectation values of the
pair products of the spherical harmonic coefficients $\langle \hat
a_{\ell m} \hat a^*_{\ell^\prime m^\prime}\rangle$ are related to the
correlation function $C(\hat q,\hat q^\prime)$ by
\begin{equation}
\langle \hat a_{\ell m}\,\hat a^*_{\ell^\prime m^\prime}\rangle = \int
d\Omega_{\hat q}\int d\Omega_{\hat q^\prime}\left\langle\dT(\hat
q)\dT(\hat q^\prime)\right\rangle Y_{\ell m}(\hat q) Y^*_{\ell^\prime
m^\prime}(\hat q^\prime) = \int d\Omega_{\hat q}\int d\Omega_{\hat
q^\prime}\,C(\hat q,\hat q^\prime) \,Y_{\ell m}(\hat q)\,
Y^*_{\ell^\prime m^\prime}(\hat q^\prime)\,. 
\label{alm_corr}
\end{equation}

In comparing theoretical predictions to observations, the power
spectrum, ${\cal C}_\ell$, is a useful ``data compression'' of
the $N(N+1)/2$ numbers in the $N\times N$ correlation matrix, $C_T$,
into $\approx N$ multipoles values of ${\cal C}_\ell$ . However, it is
sometimes overlooked (see \eg \cite{corn_sper99,inoue99b}) that a necessary
condition for this compression to be ``loss free'' is that the CMB
fluctuations are {\em statistically isotropic}, \ie the ensemble
averages such as $C(\hat q,\hat q^\prime)$ or $\langle \hat a_{\ell m}
\hat a^*_{\ell m}\rangle$ are invariant under rotation.

The standard (simply connected, FRW) universe models respect global
isotropy and predict a statistically isotropic CMB sky. The full
angular correlation function $C(\hat q,\hat q^\prime)$ between two
directions is then solely a function of their separation, $\hat q
\cdot \hat q^\prime$. Using eq.(\ref{alm_corr}), the equivalent
statement is that $\langle \hat a_{\ell m} \hat a^*_{\ell^\prime
m^\prime}\rangle = 2\pi {\cal C}_\ell/(\ell(\ell+1))
\,\delta_{\ell\ell^\prime}\, \delta_{m m^\prime}$, is diagonal in the
$(\ell,m)$ space and independent of $m$.  The angular power spectrum
${\cal C}_\ell$ is uniquely related to the correlation function
through the expansion (\ref{Ctheta_Cl}) and contains the same
information.

All Euclidean or Hyperbolic compact spaces violate global isotropy and
the CMB temperature fluctuations are {\em statistically anisotropic},
\ie $C(\hat q,\hat q^\prime)\not\equiv C(\hat q\cdot\hat
q^\prime)$. As a consequence, the $\langle \hat a_{\ell m} \hat
a^*_{\ell^\prime m^\prime}\rangle$ now define the relevant angular
power spectra, and these are are no longer required to be diagonal and
independent of $m$.  Thus, ${\cal C}_\ell$ as defined by
eq.~(\ref{def_Cl}) misses both off-diagonal pair products of $a_{\ell
m}$'s and isotropizes further by summing up the $m$-dependent diagonal
$a_{\ell m}$ pair products.

Figure~\ref{fig:cross_alm} shows the normalized full angular power spectrum 
\begin{equation}
\rho_{\ell m}^{\ell^\prime m^\prime} \equiv \langle \hat a_{\ell
m}\,\hat a^*_{\ell^\prime m^\prime}\rangle/\sqrt{\langle \hat a_{\ell
m}\,\hat a^*_{\ell m}\rangle\langle \hat a_{\ell^\prime
m^\prime}\,\hat a^*_{\ell^\prime m^\prime}\rangle}
\end{equation} 
for all the $(\ell,m)$ with $2\le\ell\le 10$ for the CH models
considered in our paper. This cross correlation coefficient clearly
shows the the existence of significant off-diagonal $\langle \hat
a_{\ell m}\,\hat a^*_{\ell^\prime m^\prime}\rangle$ in the m004(-5,1)
at $\Omega_0=0.3$ where $\rho_{\ell m}^{\ell^\prime m^\prime}$ ranges
between $-0.5$ up to $0.6$. This is the most extreme case in our set
of models since it has the smallest size of the SLS relative to the
Dirichlet domain ($V_{\rm LS}/V_{\!\cal M}=153.4$) amongst the models
discussed in this paper. As $V_{\rm LS}/V_{\!\cal M}$ decreases, the
number of $(\ell,m)$ pairs with high $\rho_{\ell m}^{\ell^\prime
m^\prime}$ decreases and the $\langle \hat a_{\ell m}\,\hat
a^*_{\ell^\prime m^\prime}\rangle$ matrix becomes closer to
diagonal. However, for the same CH space the range of $\rho_{\ell
m}^{\ell^\prime m^\prime}$ remains comparable, \eg for m004(-5,1) at
$\Omega_0=0.9$ ($V_{\rm LS}/V_{\!\cal M}=1.2$), $\rho_{\ell
m}^{\ell^\prime m^\prime}$ ranges from $-0.42$ to $0.5$ for
$2\le\ell\le 10$. The larger CH v3543(2,3) model is more isotropic,
the $\langle \hat a_{\ell m}\,\hat a^*_{\ell^\prime m^\prime}\rangle$
are somewhat more diagonal with most of the significant off-diagonal
terms at low $\ell$. The ranges in $\rho_{\ell m}^{\ell^\prime
m^\prime}$ are $\pm 0.29$ and $\pm 0.23$ for $\Omega_0=0.6$ and $0.8$,
respectively.  The $m$-dependence of the diagonal products
$\langle\hat a_{\ell m} \hat a^*_{\ell m} \rangle$ is shown in
Figure~\ref{fig:diag_alm}.

\begin{figure}
\plottwosideR{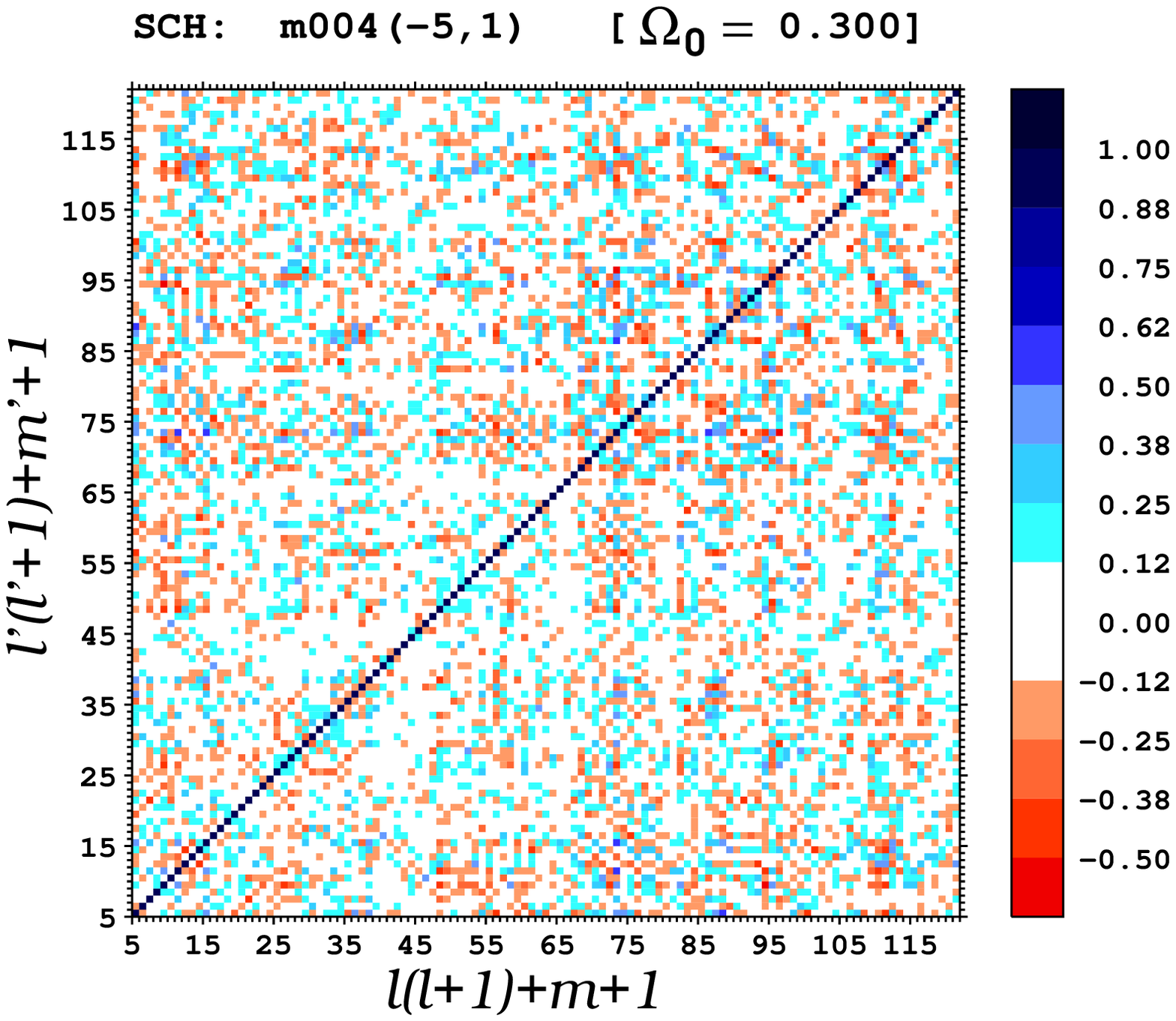}{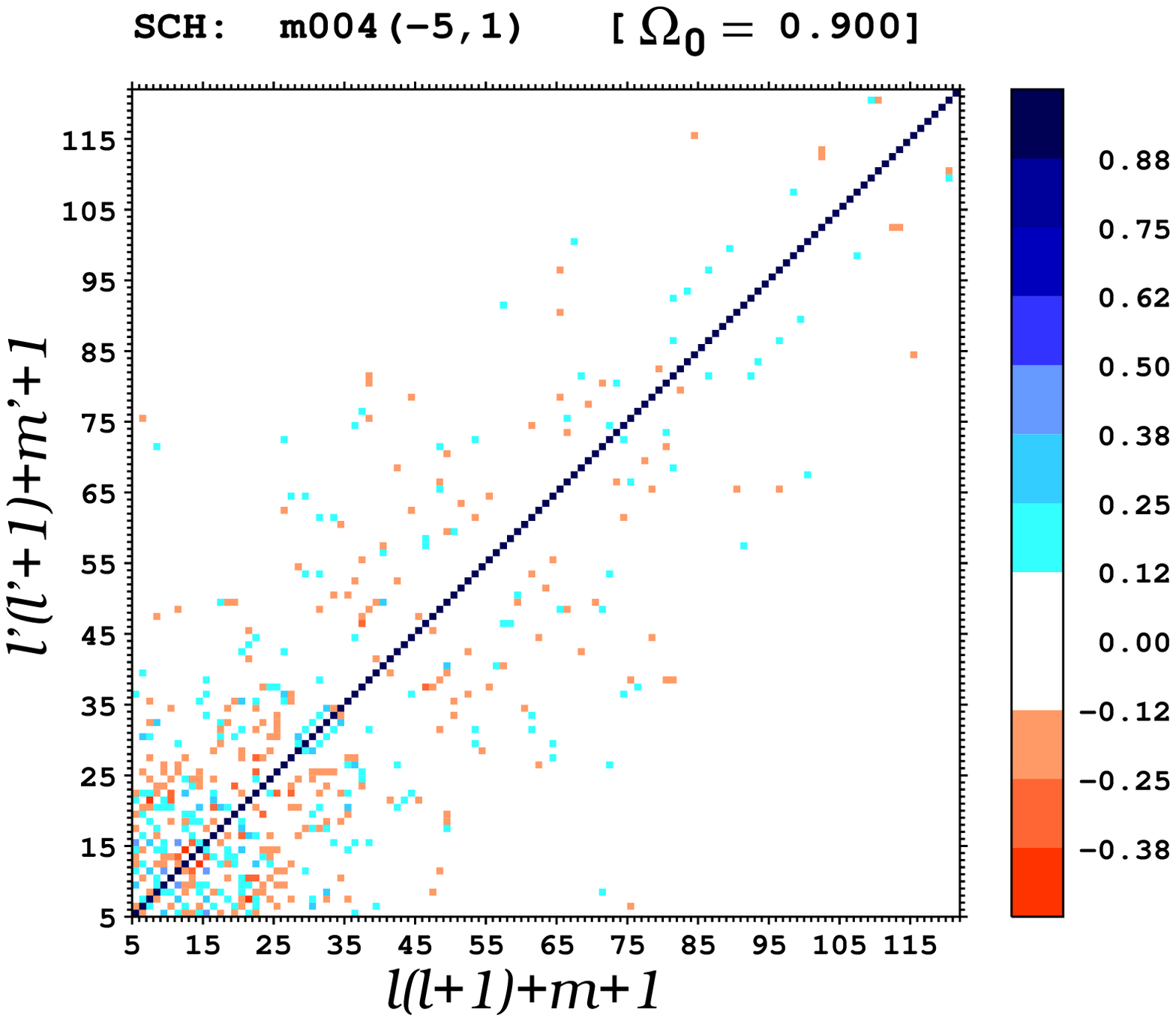}
\caption{The figure demonstrates the nondiagonal nature of the
expectation values of $a_{\ell m}$ pair products when the CMB
anisotropy is statistically anisotropic using the examples of the SCH
model at $\Omega_0=0.3$ ({\em left panel}) and $\Omega_0=0.9$ ({\em
right panel}). We use a single positive integer index, $n =
\ell(\ell+1)+m+1$ to uniquely represent each $(\ell,m)$. Each square
represents a pair of $(\ell, m)$ spherical harmonic indices for all
$(\ell,m)$ for $2\le\ell\le 10$. The cross correlation
coefficient, $\rho_{\ell m}^{\ell^\prime m^\prime} \equiv \langle \hat
a_{\ell m}\,\hat a^*_{\ell^\prime m^\prime}\rangle/\sqrt{\langle \hat
a_{\ell m}\,\hat a^*_{\ell m}\rangle\langle \hat a_{\ell^\prime
m^\prime}\,\hat a^*_{\ell^\prime m^\prime}\rangle}$ at a given $(n,
n^\prime)$ square is represented by the level of grey (or color) shown
in the accompanying palette. (The squares corresponding to pairs with
mild cross correlation, $|\rho_{\ell m}^{\ell^\prime m^\prime}|<0.12$,
have been left blank to highlight the strong ones). For statistically
isotropic CMB anisotropy only the diagonal $\rho_{\ell m}^{\ell m}=1$
terms will be nonzero. As expected, the CMB anisotropy in the SCH
model is more isotropic at $\Omega_0=0.9$ as evident from the more
diagonal nature of $\rho_{\ell m}^{\ell m}$ in the right panel.}
\label{fig:cross_alm}
\end{figure}

\begin{figure}
\plotone{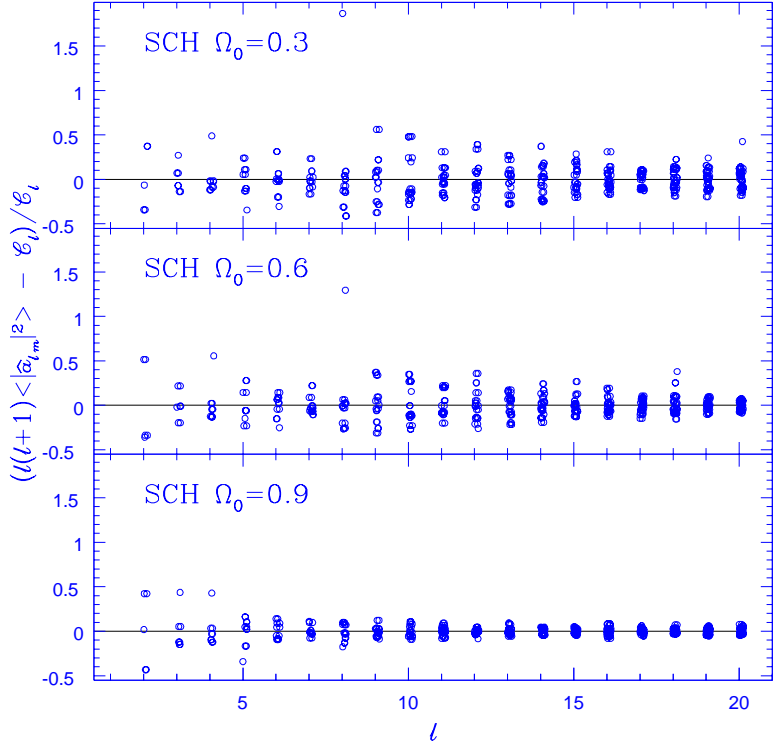}
\caption{The three panels of the figure show the relative deviation of
$\ell(\ell+1)\langle \hat a_{\ell m}\,\hat a^*_{\ell^\prime
m^\prime}\rangle$ from the mean value ${\cal C}_\ell$ for all $(\ell,
m)$ ( $2\le\ell\le20$) in the SCH model at $\Omega_0=0.3$, $0.6$ \&
$0.9$, respectively. The non zero deviations demonstrate that the
diagonal $\langle \hat a_{\ell m}\,\hat a^*_{\ell^\prime
m^\prime}\rangle$ are not independent of $m$ when the CMB is
statistically anisotropic. The relative deviations decrease with
$\Omega_m$ as the space become larger relative to the SLS. The point
with large deviation in the upper two panels corresponds to
$(\ell,m)=(8,0)$. }
\label{fig:diag_alm}
\end{figure}
 
For statistically anisotropic CMB fluctuations, the ${\cal C}_\ell$
contains less information than the full correlation matrix $C(\hat
q,\hat q^\prime)$ independent of the underlying statistics.  For
concreteness, we now focus our discussion on Gaussian random CMB
fluctuations which are completely specified by $C(\hat q,\hat
q^\prime)$. For statistically anisotropic CMB, the incompleteness of
the information contained in the ${\cal C}_\ell$ is reflected in the
enhanced cosmic variance, $\langle{\cal C}_\ell^2\rangle$.  Let us
split the correlation matrix into an isotropic part determined
by ${\cal C}_\ell$ through eq.(\ref{Ctheta_Cl}) and an anisotropic
term containing the remainder, \ie
\begin{equation}
C(\hat q,\hat q^\prime) = C^{\sc I}(\hat q,\hat q^\prime) + C^{\sc
A}(\hat q,\hat q^\prime)
\label{corr_split}
\end{equation}
so that, by definition, the isotropic part $ C^{\sc I}(\hat q,\hat
q^\prime) $ is described by the set of coefficients $C_l$ of Legendre
series
\footnote{For simplicity we ignore the experimental window function, $
W(\hat q,\hat q^\prime)$. For \cobedmr it is isotropic, with $W_\ell
\equiv B^2_\ell $, (where
$B_\ell$ is spherical transform of the isotropic experimental beam,)
and the effect of including the window function is to scale the ${\cal
C}_\ell$ by $B^2_\ell$.}

\begin{equation}
C^{\sc I}(\hat q,\hat q^\prime) = \sum_{\ell} \frac{\ell
+1/2}{\ell(\ell +1)}~{\cal C}_\ell P_\ell(\hat q\cdot\hat q^\prime)\, ,
\label{defisocorr}
\end{equation}
and the anisotropic part $ C^{\sc A}(\hat q,\hat q^\prime)$ is
orthogonal to the Legendre polynomials
\begin{equation}
\int d\Omega_{\hat q}\int d\Omega_{\hat q^\prime}\, C^{\sc A}(\hat q,\hat
q^\prime)\, P_\ell(\hat q\cdot\hat q^\prime) = 0\,.
\label{defanisocorr}
\end{equation}

The presence of the nonzero anisotropic part $ C^{\sc A}(\hat q,\hat
q^\prime) $ is the main attribute of statistically anisotropic models,
and in particular of the CH models.  Consider now the effect this term
has on the probability distribution $P(\tilde {\cal C}_\ell)$ of the
isotropized power spectrum estimator
\begin{equation}
\tilde {\cal C}_\ell \equiv \frac{\ell(\ell+1)}{2\pi (2\ell +1)}~
\sum_{m=-\ell}^{\ell} \int d\Omega_{\hat q}\int d\Omega_{\hat
q^\prime}\dT(\hat q)\dT(\hat q^\prime) Y_{\ell m}(\hat q) Y^*_{\ell
m}(\hat q^\prime) \, .
\end{equation}
Of course, the expectation value of the estimator 
is determined solely by $ C^{\sc I}(\hat q,\hat q^\prime)$ (as it
should be using eq.(\ref{defanisocorr})): 
\begin{equation}
\langle\tilde {\cal C}_\ell\rangle =
\frac{\ell(\ell+1)}{8\pi^2}\int d\Omega_{\hat q}\int
d\Omega_{\hat q^\prime}\, C(\hat q,\hat q^\prime)\, P_\ell(\hat
q\cdot\hat q^\prime) \, .
\label{clmean}
\end{equation}
However, the distribution $P(\tilde {\cal C}_\ell)$ depends on the anisotropic part $ C^{\sc
A}(\hat q,\hat q^\prime) $. In particular, the variance of $\tilde
{\cal C}_\ell$, which can be calculated as a four-point correlation of the $\Delta T$'s, 
is enhanced due to the influence of $ C^{\sc A}(\hat q,\hat q^\prime) $:
\begin{eqnarray}
{\rm var}(\tilde {\cal C}_\ell)\equiv \langle \tilde {\cal C}_\ell^2\rangle -
\langle\tilde{\cal C}_\ell\rangle^2 &=& 2 \left[\frac{\ell(\ell+1)}{8\pi^2}
\right]^2 \int d\Omega_{\hat q_1}\int d\Omega_{\hat
q_2}\int d\Omega_{\hat q_3}\int d\Omega_{\hat q_4}\, C(\hat q_1,\hat
q_3) P_\ell(\hat q_1\cdot\hat q_2)\, C(\hat q_2,\hat q_4)\,
P_\ell(\hat q_3\cdot\hat q_4) \nonumber\\ &=&  \frac{2 \langle \tilde {\cal C}_\ell \rangle^2}{2\ell +1}
+ \frac{\ell^2(\ell+1)^2}{32\pi^4} \int d\Omega_{\hat q_1}\int
d\Omega_{\hat q_2}\left[ \int d\Omega_{\hat q_3}\,
C^{\sc A}(\hat q_1,\hat q_3) \,P_\ell(\hat q_2\cdot\hat q_3) \right]^2\,.
\label{clvar}
\end{eqnarray}
In the final expression, the first term is the well known result for
the cosmic variance of ${\cal C}_\ell$, strictly valid only for
statistically isotropic CMB fluctuations. The second term was obtained
using the fact that $ C^{\sc A}(\hat q_1,\hat q_2) $ and $P_\ell(\hat
q_1\cdot\hat q_2)$ are symmetric functions.  It represents a positive
definite correction to the standard cosmic variance arising from the
anisotropy of $C(\hat q,\hat q^\prime)$. Hence, the cosmic variance is
always larger for a statistically anisotropic CMB compared to that
expected for the statistically isotropic case with the same ${\cal
C}_\ell$. 

To calculate eq.(\ref{clvar}) numerically, we took our
$C_{Tpp^\prime}$ and used matrix representations of the $P_\ell$ at
the relevant COBE-pixel pairs. As a check of accuracy, the same
procedure was applied to the infinite statistically isotropic models
and the first term in eq.(\ref{clvar}) was recovered. This is a
stringent test of the cancellations required to obtain this result. We
found that for the finite compact models as well as the infinite
models, our results are accurate for angular scales bigger than the
beam, including the finite pixel effects.

Figure~\ref{fig:clCH} shows the angular power spectrum for a set of
the CH models with the associated cosmic variance. It demonstrates
that the cosmic variance of ${\cal C}_{\ell}$ in CH models is
significantly higher than one may naively assign assuming statistical
isotropy.  As a result, ${\cal C}_{\ell}$'s do not strongly
distinguish CH models from the corresponding standard isotropic models
and, on their own, are not very restrictive.  Comparing CMB
predictions of CH spaces to data using ${\cal C}_\ell$ alone is {\em
not incorrect, but inadequate} since one has not used the full
information available.  The larger cosmic variance simply implies that
the theoretical prediction is weaker. The argument that a comparison
based on more information is more relevant is quite obvious. Any
evaluation of the relative likelihood of a model based on ${\cal
C}_\ell$ is superseded by a likelihood analysis that uses the complete
correlation information.  {\em A model that does well solely in terms
of ${\cal C}_\ell$ but fails in terms of the correlation matrix
$C_{Tpp^\prime}$ should be considered ruled out.}

In Table~\ref{tab:isologprob} we compare the relative likelihood for
our models obtained using ${\cal C}_\ell$ versus that obtained using
the full correlation information. 

The procedure we use to determine the first line in the Table is to
take the compact model's $C_{Tpp^\prime}$, use the $P_\ell$ matrices
to calculate the theoretical ${\cal C}_\ell$, as in eq.(\ref{clmean}),
then assume that we have an infinite model with exactly that power
spectrum, so that these ${\cal C}_\ell$'s encode all the information
in the theory. We then use these to calculate new $C_{Tpp^\prime}$ for
this theory, and determine the full Bayesian likelihood for it
relative to the true infinite open model with the same $\Omega_0$ and
its corresponding ${\cal C}_\ell$.  The results show that at roughly
the $+1\sigma$ level, these power spectra are preferred. Note that the
likelihood ratios in row 1 are independent of orientation. However,
the second row of the Table shows that when the full information on
angular patterns is included in the analysis, the likelihoods change
dramatically, strongly disfavouring small compact models, as in
Table~\ref{tab:logprob}. As well, the case for the $\Omega_0=0.8$ LCH
model at the best orientation is enhanced by the inclusion of the full
pattern information.

Models which fare very poorly with respect to a full correlation
comparison may well look favored based on the ${\cal C}_\ell$. The
reason for this is not hard to understand. If the compact space is not
much larger than the SLS then it predicts strong anisotropic
correlation features in the CMB sky which are at odds with the data
(see $\S$~\ref{sec_cmbcorr}). However, an isotropized measure such as
the ${\cal C}_\ell$ is insensitive to these features. This implies
that the comparison of CMB anisotropy in CH models using ${\cal
C}_\ell$ alone is grossly inadequate and could be quite misleading.

\end{document}